\documentclass[apj]{emulateapj}
\usepackage{amsmath}
\usepackage{apjfonts}

\newcommand{\DDir}{\relax{D\kern-.7em{/}}}

\newcommand{\inv}[1]{\frac{1}{#1}}

\newcommand{\ra}{\rightarrow}

\newcommand{\X}{\times}

\newcommand{\bB}{\textbf{B}}
\newcommand{\bE}{\textbf{E}}
\newcommand{\bj}{\textbf{j}}
\newcommand{\bx}{\textbf{x}}

\newcommand{\bp}{\textbf{p}}
\newcommand{\bbv}{\textbf{v}}

\newcommand{\bu}{\textbf{u}}
\newcommand{\bk}{\mathbf{k}}
\newcommand{\cbB}{\check\bB}

\newcommand{\cbE}{\check\bE}
\newcommand{\cbj}{\check\bj}

\newcommand{\cbx}{\check\bx}
\newcommand{\cbp}{\check\bp}

\newcommand{\ct}{\check t}
\newcommand{\cf}{\check f}
\newcommand{\cp}{\check p}

\newcommand{\cna}{\check\nabla}
\newcommand{\cD}{\check D}

\newcommand{\Dt}{\Delta t}
\newcommand{\Dx}{\Delta x}
\newcommand{\Dbx}{\mathbf{\Delta x}}
\newcommand{\hbz}{\mathbf{\hat z}}

\newcommand{\hbp}{\hat \bp}

\newcommand{\be}{\begin{equation}}
\newcommand{\ee}{\end{equation}}
\newcommand{\bea}{\begin{equation*}}
\newcommand{\eea}{\end{equation*}}

\newcommand{\ave}[1]{\left\langle #1\right\rangle}

\newcommand{\pr}{\partial}

\newcommand{\nin}{\relax{\in\kern-.8em{/}}}

\newcommand{\te}{\theta}
\newcommand{\al}{\alpha}

\newcommand{\de}{\delta}
\newcommand{\De}{\Delta}

\newcommand{\om}{\omega}
\newcommand{\sig}{\sigma}

\newcommand{\ep}{\epsilon}

\newcommand{\trm}{\textrm}
\newcommand{\vs}{\textrm{v}_s}
\newcommand{\vd}{\textrm{v}_d}
\newcommand{\cm}{\mbox{ cm}}

\newcommand{\se}{\mbox{ s}}

\newcommand{\km}{\mbox{ km}}

\newcommand{\muG}{\mbox{ $\mu$G}}

\begin{document}
\title{Self-Similar Collisionless Shocks}

\author{Boaz Katz\altaffilmark{1}, Uri Keshet\altaffilmark{2} and Eli
Waxman\altaffilmark{1}}

\altaffiltext{1}{Physics Faculty, Weizmann Institute, Rehovot 76100, Israel; boazka@wizemail.weizmann.ac.il, waxman@wicc.weizmann.ac.il}
\altaffiltext{2}{Institute for Advanced Study, Einstein Drive, Princeton, NJ 08540, USA; Friends
of the Institute for Advanced Study member; keshet@sns.ias.edu}

\begin{abstract}

Observations of $\gamma$-ray burst afterglows suggest that the correlation length of magnetic field fluctuations in the
downstream of relativistic non-magnetized collisionless shocks grows with the distance from the shock to scales much
larger than the plasma skin depth. We argue that this indicates that the plasma properties are described by a
self-similar solution, and derive constraints on the scaling properties of the solution. For example, we find that the
scaling of the characteristic magnetic field amplitude with distance from the shock is $B\propto D^{s_B}$ with
$-1<s_B\leq 0$, that the spectrum of accelerated particles is $dn/dE\propto E^{-2/(s_B+1)}$, and that the scaling of the
magnetic correlation function is $\ave{B_i(\bx)B_j(\bx+\Dbx)}\propto\Dbx^{2s_B}$ (for $\Dx\gg D$). We show that the
plasma may be approximately described as a combination of two self-similar components: a kinetic component of energetic
particles, and an MHD-like component representing "thermal" particles. We argue that the thermal component may be
considered as an infinitely conducting fluid, in which case $s_B=0$ and the scalings are completely determined (e.g.
$dn/dE\propto E^{-2}$ and $B\propto D^0$, with possible logarithmic corrections). Similar claims can be made regarding
non-relativistic shocks such as in supernova remnants, if the upstream magnetic field can be neglected. If
self-similarity holds, it has important implications for any model of particle acceleration and/or field generation. For
example, we show that the velocity-angle diffusion coefficient in diffusive shock acceleration models must satisfy
$D_{\mu\mu}(\bp,D)=D^{-1}\tilde{D}_{\mu\mu}(\bp/D)$ (where $\bp$ is the particle momentum), and that a previously
suggested model for the generation of a large scale magnetic field through hierarchical merger of current-filaments
should be generalized. The generalization leads to qualitative changes in the model's major predictions, e.g. a constant
merger velocity is predicted rather than increasing velocity approaching the speed of light. Finally, we point out that
the self-similarity assumptions may be tested through the implied evolution of homogenous (time-dependent) plasmas, which
may be accessible to direct numerical simulations: We predict that the inclusion (at the initial conditions) of a
power-law spectrum of high energy particles, $dn/dE\propto E^{2+l_p}$, would lead to magnetic field evolution following
$B\propto t^{-(l_p+4)/2(l_p+3)}$.

\end{abstract}
\keywords{acceleration of particles --- shock waves --- gamma rays: bursts --- magnetic fields --- plasmas --- supernova remnants}

\section{Introduction}
\label{Introduction}

Due to the low densities characteristic of a wide range of astrophysical environments, shocks observed in many astrophysical systems are collisionless, i.e. mediated by collective plasma instabilities rather than by particle-particle collisions. For example, collisionless shocks play an important role in supernova remnants \citep[e.g.][]{Blandford87}, jets of radio galaxies \citep[e.g.][]{Begelman94,Maraschi03}, gamma-ray bursts \citep[GRB's, e.g.][]{Zhang04}, and the formation of the large scale structure of the Universe \citep[e.g.][]{Loeb00}. Although collisionless shocks have been studied for several decades, theoretically and experimentally, in space and in the laboratory, a self-consistent theory of collisionless shocks based on first principles has not yet emerged \citep[see, e.g., comments in][]{Krall97}.

Observations of GRB "afterglows," the delayed low energy emission
following the prompt $\gamma$-ray emission, provide a unique probe
of the physics of collisionless shocks. Current understanding
suggests that the afterglow radiation observed is the synchrotron
emission of energetic, non-thermal electrons in the downstream of
a strong collisionless shock driven into the surrounding
interstellar medium (ISM) or stellar wind. These collisionless
shocks start out highly relativistic, with shock Lorentz factor
$\gamma_s\sim 100$ on time scale of minutes after the GRB, and
gradually decelerate to $\gamma_s\sim10$ on a day time scale and
$\gamma_s\sim1$ on a month time scale. This allows one to probe
the physics of the shocks over a wide range of Lorentz factors.
Afterglow shocks are highly "non-magnetized:" The ratio of
magnetic field to kinetic energy flux ahead of the shock is very
small, $U_{B1}/n_1m_pc^2\sim 10^{-9}$, where $U_{B1}$ and $n_1$
are the magnetic energy density and particle number density in the
upstream rest-frame. This strongly suggests that the shock
structure is determined by the upstream density and the shock
Lorentz factor alone \citep[e.g.][]{Gruzinov01a}. We therefore
adopt the assumption that the shock structure approaches a well
defined limit as $U_{B1}/n_1m_pc^2\rightarrow0$, and that GRB
afterglow shocks are approximately described by this limiting
solution (see \S~\ref{sec:AreThere} for a detailed discussion).

It may be noted here that the upstream density can be eliminated
from the problem by measuring time in units of the (shock-frame
proton) plasma time, $\omega_p^{-1}=(4\pi\gamma_s n_1 e^2 / m_p
\gamma_s)^{-1/2}=(4\pi n_1 e^2 / m_p)^{-1/2}$, and measuring
distances in units of the corresponding skin depth,
$l_{sd}=c/\omega_p$. The shock is then completely specified by the
dimensionless parameter $\gamma_s\vs/c$ (and the dimensionless
mass ratio $m_e/m_p$; the upstream pressure is assumed
negligible). In this sense, GRB shocks may be considered "simple."

The synchrotron model of GRB afterglows requires a strong magnetic field and a population of energetic electrons to be present in the downstream. Optical observations \citep[e.g.][]{Zhang04}, the clustering of explosion energies \citep{Frail01}, and the observed X-ray luminosity \citep{Freedman01,Berger03} suggest that the fraction of post-shock thermal energy density carried by non-thermal electrons, $\epsilon_e$, is large, $\epsilon_e\approx0.1$. The fraction of post-shock thermal energy carried by the magnetic field, $\epsilon_B$, is less well constrained by observations. However, in cases where $\epsilon_B$ can be
reliably constrained by multi waveband spectra, values close to equipartition, $\epsilon_B\sim 0.01$ to $0.1$, are inferred \citep[e.g.][]{FWK00}\footnote{\citet{Eichler05} have pointed out that observations determine $\epsilon_e$ and $\epsilon_B$ only up to a factor $f$, the fraction of electrons accelerated, where $m_e/m_p<f<1$. However, it is expected that $f$ is not very small, $f\gtrsim 1/10$ \citep{Eichler05}.}.

Near equipartition magnetic fields may conceivably be produced in the collisionless shock driven by the GRB explosion by
electromagnetic (e.g. Weibel-like) instabilities \citep[e.g.][]{Blandford87,Gruzinov99,Medvedev99,Wiersma04}\footnote{It
should be noted here that if a hydrodynamic description of the shock were applicable, the shock would have been stable
and no magnetic fields could have been generated \citep{Gruzinov00,Wang02}. A possible caveat is the presence of large
inhomogeneities in the upstream plasma (ISM), which may drive turbulence in the downstream plasma, which may generate
strong magnetic fields with large correlation length}. The main challenge associated with the downstream magnetic field
is related to the fact that in order to account for the observed radiation as synchrotron emission from accelerated
electrons, the field amplitude must remain close to equipartition deep into the downstream, over distances
$\sim10^{10}l_{sd}$: At $t\sim1$~d the magnetic field must be strong throughout the (proper) width $\Delta\sim2\gamma_s c
t\sim10^{17}$~cm while $l_{sd}\sim10^7(n_1/1{\rm cm^{-3}})^{-1/2}$~cm. This is a challenge since electromagnetic
instabilities are believed to generate (near-equipartition) magnetic fields with coherence length $L\sim l_{sd}$, and a
field varying on such scale is expected to decay within a few skin-depths downstream \citep{Gruzinov01a}. This suggests
that the correlation length of the magnetic field far downstream must be much larger than the skin depth, $L\gg l_{sd}$,
perhaps even of the order of the distance from the shock \citep{Gruzinov99,Gruzinov01a}.

Growth of the characteristic length scale by many orders of magnitude, from $L\sim l_{sd}$ to $L\gg l_{sd}$, is a strong
indication of self-similarity. In regions where $L\gg l_{sd}$, which also imply that $L$ is much larger than the Larmor
radius of thermal protons ($L\gg R_{L,th}\sim \gamma_sm_pc^2/eB\sim l_{sd}\ep_B^{-1/2}$), it is reasonable to assume that
$L$ is the only relevant length scale, and self-similarity is expected (see \S~\ref{sec:self-sim} for a detailed
discussion). The main goal of the current paper is to introduce and formulate the assumption of a self-similar
collisionless shock structure, and to study some of its consequences.

Although our analysis is motivated by GRB afterglow observations,
it may be relevant also for non-relativistic collisionless shocks,
such as shocks in young supernova remnants (SNRs). In the past few
years, high resolution X-ray observations have provided indirect
evidence for the presence of strong magnetic fields, $\gtrsim
100\muG$, in the \emph{non-relativistic}
($\textrm{v}_s\sim\mbox{few }\times1000\km\se^{-1}$) shocks of
young SNRs (see \citealt{Bamba03, Vink03, Volk05}). These fields
extend to distances $D>10^{17}\cm \sim 10^{10} l_{sd}$ downstream,
and possibly even $\gtrsim 10^{16}\cm$ upstream, of the shock. In
resemblance to GRB's, such strong magnetic fields cannot result
from the shock compression of a typical interstellar medium (ISM)
magnetic field, $B_1\sim \mbox{few }\muG$. In SNRs, the
discrepancy is somewhat less severe, $U_{B1}/n_1 m_p\vs^2\lesssim
10^{-4}$, and the possibility that these magnetic fields are
related to the large scale ISM fields cannot be ruled out. If the
ISM magnetic fields can be neglected, this suggests that these
shocks too may have a self-similar nature. Henceforth, when
discussing non-relativistic shocks, we assume that this is indeed
the case.

The non-thermal energetic electron (and proton) population is believed to be produced by the diffusive (Fermi) shock
acceleration (DSA) mechanism (for reviews see \citealt{Drury83,Blandford87,Malkov01}). Acceleration of charged particles
to high, non-thermal energies is a ubiquitous phenomenon in both relativistic and non-relativistic collisionless shocks.
The accelerated particles are estimated to carry a considerable part of the energy: electrons alone carry $\sim 10\%$ of
the thermal energy in GRB external shocks \citep{Eichler05} and $\sim 5\%$ of the thermal energy in SNR shocks
\citep[][and references therein]{Keshet04}; and at least $10\%$ of the energy in SNR shocks must be converted into
relativistic protons if these shocks are responsible for Galactic cosmic rays \citep{Drury89}. This has several important
implications. The accelerated particles are likely to have an important role in generating and maintaining the inferred
magnetic fields. This conclusion is supported also by the evidence of strong amplification of the magnetic field in the
upstream of GRB afterglow shocks \citep{Zhuo06}, which is most likely due to the streaming of high energy particles ahead
of the shock. Since the high energy particles are likely to play an important role in the generation of the fields, a
theory of collisionless shocks must provide a self-consistent description of particle acceleration, which depends on the
scattering of these particles by magnetic fields, and field generation, which is likely driven by the accelerated
particles.

The search for a self-consistent theory of collisionless shocks has led to extensive numerical studies. Particle in cell
(PIC) simulations were performed in one dimension (e.g. \citealt{Dieckmann06}), in two dimensions (2D; e.g.
\citealt{Wallace91,Kato05}), in two spatial and three velocity dimensions (2D3V; e.g.
\citealt{Gruzinov01a,Gruzinov01b,Medvedev05}) and in the past few years also in three dimensions (3D; e.g.
\citealt{Silva03,Nishikawa03,Frederiksen04,Jaroschek04,Spitkovsky05}). Such simulations have provided compelling evidence
that transverse, electromagnetic (Weibel-like) instabilities generate near-equipartition magnetic fields in pair
($e^+e^-$) plasma, and $\epsilon_B\gtrsim m_e/m_p$ magnetic fields in ion-electron plasma. However, 3D simulations are
limited to very small simulation boxes, and at present can reliably probe small length scales no larger than $\sim 100$
electron skin-depths, and short time scales no longer than $\sim 100$ electron plasma times. Hence, there is only
preliminary evidence for the existence of collisionless shocks in 3D, and only in a pair-plasma (Simulations of
ion-electron plasma are forced to employ an effective, small proton to electron mass ratio, $\tilde{m}_p/m_e\lesssim 20$
with present computational resources, and the preliminary results thus obtained are not easily extrapolated to more
realistic mass ratios). Obviously, the question of field survival and correlation length evolution on length scales $\gg
l_{sd}$ are not yet answered. Similarly, highly energetic particles cannot be contained in the small simulation boxes
used, so Fermi-like acceleration processes are suppressed. It is important to note, that some published results are based
on PIC simulations in stages where the boundary conditions strongly modify the plasma evolution. For example, claims that
the magnetic fields decay slowly or saturate at some finite level remain questionable, until verified by simulations with
sufficiently large simulation boxes. A discussion of 3D PIC simulations and their physical implication appears in
Appendix \S~\ref{appendix:PICs}.

In \S~\ref{sec:NMShock} we lay the basis for our analysis of
afterglow (relativistic) and SNR (non-relativistic) shocks. In
\S~\ref{sec:AreThere} we discuss our assumption that these shocks
 are highly "non-magnetized," i.e. that the shock structure
approaches a well defined limit as $U_{B1}/n_1m_p\vs^2\rightarrow0$ and that the shocks observed are approximately
described by this limiting solution. In \S~\ref{sec:EOM} we present the governing equations and discuss their dependence
on dimensional parameters, which is relevant for the discussion of self-similarity. In particular, we demonstrate that
when distances are measured in units of $l_{sd}$, the shock structure depends only on $\gamma_s\vs/c$ (and $m_e/m_p$). In
\S~\ref{sec:stationary} we clarify the notion of "shock structure:" Since the electromagnetic fields and particle
distributions fluctuate with time (at any given point downstream, and perhaps also upstream of the shock), the
"stationary shock structure" is given by the correlation functions of the fluctuating quantities (which are expected to
depend only on the distance from the shock). We define a notation that is useful for the discussion of the
self-similarity assumption.

In \S~\ref{sec:self-sim} we introduce and formulate the assumption of a self-similar structure in the downstream of
non-magnetized collisionless shocks, derive several scaling relations of the physical quantities, and discuss some of the
physical implications. One of the conclusions of \S~\ref{sec:self-sim} is that the plasma may be approximately described
as a combination of two self-similar components: a kinetic component of energetic particles, and an MHD-like component
representing the bulk, "thermal" particles. The MHD-like component is discussed in \S~\ref{sec:fluid}. We argue that this
component may be treated as an infinitely conducting fluid, and show that this leads to a complete determination of the
scaling laws.

In \S~\ref{sec:OtherCases} we present various extensions of the analysis. Self-similarity is studied in the upstream of
non-magnetized collisionless shocks and in homogenous time-dependent plasmas, which may be more accessible to simulations
than (non-homogeneous) collisionless shocks. In \S~\ref{sec:general} we discuss some of the implications of the
self-similarity assumption to models of diffusive particle acceleration, and to the phenomenological model, suggested by
\citet{Medvedev05}, of field generation through hierarchical merger of electric current filaments. The latter model is
generalized, and shown to follow the self-similar scalings. Our main results and conclusions are summarized in
\S~\ref{sec:Discussion}.

\section{Non-magnetized collisionless shocks}
\label{sec:NMShock}

\subsection{Do non-magnetized collisionless shocks exist?}
\label{sec:AreThere}

In general, the parameters determining the structure of a collisionless shock (for a given homogenous upstream composition which we assume to be an electron-proton plasma) are the shock (four) velocity $\gamma_s\vs$, and the upstream density $n_1$, pressure $P_1$ and magnetic field $\bB_1$. We can formally ask what the limiting configuration is when
$\bB_1$ and $P_1$ approach zero. We assume that in this limit there is some shock solution, to which we refer as
non-magnetized and strong. It is important to stress that by "non-magnetized" ("strong") we refer to the null effect of
the value of the upstream magnetic field (pressure) on the stationary configuration of the shock, regardless of the
dynamical role of these parameters in the generation of the shock.

We next ask how small should $\bB_1$ and $P_1$ be so that the solution is no longer affected by their value.
The dimensionless parameters $\sig_B=U_{B1}/n_1m_p\vs^2,\sig_P=P_1/n_1m_p\vs^2$ measure the relative contribution of the
magnetic field and thermal energy to the energy flux in the shock frame (which is constant and thus equal to its value in
the far upstream),
\begin{align}
&T^{0z}\sim\gamma_s^2n_1m_p\vs^2(1+\sig_B+\sig_P)\vs,
\end{align}
where we assumed that the shock normal (the direction of the flow) is in the $\hbz$ direction. It is reasonable to assume
that if $\sig_B$ and $\sig_P$ are much smaller than unity, the magnetic field and pressure do not affect the shock
structure. Since $\sig_B\sim(l_{sd}/R_{L,th})^2$, the condition $\sig_B\ll 1$ implies that the thermal protons in the
shock frame are not affected by the compressed upstream magnetic field on the dynamical distance $\sim l_{sd}$ (the
cyclotron time $\sim \gamma_s m_pc/eB$ is longer than the dynamical time $\sim \om_p^{-1}$ as long as $\sig_B<
(c/\vs)^2)$.

Simulations of relativistic shocks with various values of the upstream magnetic field (for a shock with Lorentz factor
$\Gamma=30$ in electron-positron plasma) are reported in \citet{Spitkovsky05}, and a typical value of $\sig_B\sim
10^{-2}$ is found to distinguish between the regimes of magnetized and non-magnetized shocks. Note, that this result may
change considerably for electron-ion plasmas or if a population of accelerated particles is included, and may be affected
by the transverse (perpendicular to the flow) boundary conditions (see \S~\ref{appendix:PICs}). There is some debate
regarding our assumption that the magnetic field in the upstream of GRB external shocks is insignificant.
\citet{Lyubarsky05} claim that the initial stages of the generation of the shock depend on the upstream magnetic field,
because it can isotropize the protons faster than the magnetic fields generated in the shock can. Note, however, that
even if the upstream magnetic field does play an important dynamical role in the evolution of the shock, this does not
imply that the steady-state solution depends on the exact value of the upstream magnetic field.

Similarly, \citet{Bell04,Bell05} and \citet{Milosavljevic05b} claim that the cosmic ray precursor of the shock can
amplify the upstream magnetic field significantly. Assuming that the magnetic field is amplified by many orders of
magnitude, it is reasonable to assume that its initial value is not important.

\subsection{Governing equations and dimensional considerations}
\label{sec:EOM}

Henceforth, we assume that strong, non-magnetized collisionless
shocks do indeed exist. We consider a quasi steady-state planar shock. First, we consider the equations that govern the
full plasma distribution function, $f_\al(\bx,\bp,t)$, and are valid globally, i.e. both upstream and downstream.
Subscripts $\al\in\{e,i\}$ denote electrons and ions, respectively. Approximate equations, which are valid only in the
far downstream and for which self-similarity is assumed, are discussed in \S~\ref{sec:2comp} and \S~\ref{sec:fluidEqs}.

The flow is governed by Vlasov's equation,
\begin{equation}\label{eq:Vlasov}
\pr_tf_\al+\bbv(\bp)\cdot\nabla f_\al+q_\al\left(\bE+\inv{c}\bbv\X\bB\right)\cdot\nabla_{\bp}f_{\al}=0,\end{equation} and
Maxwell's equations,
\begin{align}\label{eq:Maxwell}
&\nabla\X \bB=\frac{4\pi}{c}\bj+\inv{c}\pr_t\bE,\cr
&\nabla\X\bE=-\inv{c}\pr_t\bB,
\end{align}
where the electric current is related to $f_\al$ through
\begin{align}\label{eq:Currentrelation1}
&\bj=\sum_\al q_\al\int d^3\bp_\al \bbv(\bp)f_\al(\bp).
\end{align}
The velocity of the particles is given in terms of the momentum by
\begin{equation}
\bbv(\bp)=c\frac{\bp}{\sqrt{m_{\al}^2c^2+p^2}}.
\end{equation}
As usual, the equations $\nabla\cdot \bB=0$ and $\nabla\cdot\bE=4\pi\rho$ are assumed to hold
at some initial time, and are therefor preserved by the other equations at all times.

The shock is completely defined by the far upstream boundary conditions. Written in the shock frame, these conditions are
\begin{align}
&\bE(D\ra-\infty)=\bB(D\ra-\infty)=0,\cr
&f_\al(D\ra-\infty)=\gamma_s n_1\de^3(\bp-\gamma_s\vs m_\al \hbz),
\end{align}
where $D$ is the distance from the shock front in the direction of the downstream ($D$ is negative in the upstream).
Note that the equations are independent of the frame of reference.

In order to highlight the dimensional dependencies of the Vlasov-Maxwell equations, we express the physical quantities
$A=t,\bx\ldots$ in term of dimensionless variables $\check A$,
\begin{align}\label{eq:DimensionlessQuantities}
&t=\frac{l_{sd}}{\vs}\ct,~~\bx=l_{sd}\cbx
,~~\bp=\frac{m_\al}{m_p}p_{th}\cbp,\cr
&\bB=\frac{p_{th}c}{el_{sd}}\cbB,~~
\bE=\frac{p_{th}\vs}{el_{sd}}\cbE,\cr &\bj=e\vs\gamma_sn_1\cbj,
\end{align}
and
\begin{align}\label{eq:DimensionlessQuantities2}
&f_\al=\frac{\gamma_sn_1}{\left(\frac{m_\al}{m_p}p_{th}\right)^3}\cf_\al,\cr
\end{align}
where
\begin{align}\label{eq:DimensionlessQuantities3}
&\om_{pi}^2=\frac{4\pi n_1 e^2}{m_p}\sim\frac{4\pi n_2 e^2}{\gamma_s m_p} ,\cr &l_{sd}=\frac c{\om_{pi}},\cr
&p_{th}=m_p\gamma_s\vs,\cr
\end{align}
are the characteristic ion plasma frequency (squared), ion skin-depth and characteristic momentum of thermal particles in
the downstream. Inserting these into Eqs. \eqref{eq:Vlasov} and \eqref{eq:Maxwell}, we obtain
\begin{align}\label{eq:DimensionlessEq}
\pr_{\ct}\cf_\al+\frac{\bbv}{\vs}\cdot\cna\cf_\al+&\frac{q_\al/e}{m_\al/m_p}\left(\cbE+\frac{\bbv}{\vs}\X\cbB\right)\cdot\nabla_{\cp}\cf_\al=0,\cr
&\cna\X\cbB=\cbj+\frac{\vs^2}{c^2}\pr_{\ct}\cbE,\cr
&\cna\X\cbE=-\pr_{\ct}\cbB
\end{align}
and
\begin{align}\label{eq:DimensionlessCurrent}
&\cbj=\sum_\al\frac{q_\al}{e}\int d^3\cbp\frac{\bbv}{\vs}\cf_\al,
\end{align}
with
\begin{align}\label{eq:DimensionlessVelocity}
&\frac{\bbv}{\vs}=\frac{\cbp}{\sqrt{\inv{\gamma_s^2}+\frac{\vs^2}{c^2}\cp^2}}.\cr
\end{align}

The boundary conditions at upstream infinity (written in the shock frame) can similarly be written in dimensionless form,
\begin{align}\label{eq:DimensionlessBc}
&\cbE(\cD\ra-\infty)=\cbB(\cD\ra-\infty)=0\end{align} and
\begin{align}\label{eq:DimensionlessBc2}
\cf_\al(\cD\ra-\infty)=\de^3(\cbp-\hbz),\cr
\end{align}
where $\cD=D/l_{sd}$.

Measuring distances in units of $l_{sd}$, we have thus arrived at a set of dimensionless equations,
Eqs.~\eqref{eq:DimensionlessEq}-\eqref{eq:DimensionlessBc2}, which depend only on $\gamma_s\vs/c$ (and on $m_e/m_p$).

\subsection{Stationary shock structure}
\label{sec:stationary}

The electromagnetic fields and particle distributions at any given point behind (downstream of) the shock fluctuate with
time. For each set of values of the shock parameters, $n_1$ and $\gamma_s\vs$, there is a large ensemble of
time-dependent "specific solutions," with specific temporal and spatial dependence of particle distribution functions and
electromagnetic fields. We assume that the averages and correlation functions of the fluctuating quantities depend only
on the distance from the shock, and are identical for all specific solutions (in the limit that the size of the shock
plane is infinite).

Consider for example the particle distribution function $f_\al(\bx=D\hbz+\bx_\perp,\bp,t)$ or magnetic field
$\bB(\bx=D\hbz+\bx_\perp,t)$, where we have separated the $\bx$ dependence to dependence on the distance from the shock
front, using $\hbz$ as the direction of the shock normal, and on $\bx_\perp$, a two dimensional vector perpendicular to
$\hbz$. While $f_\al$ depends on $t$ and $\bx_\perp$, we assume that an average of $f_\al$ over planes perpendicular to
the shock normal
is independent of $t$ and $\bx_\perp$:
\begin{align}\label{eq:Avaregedf}
\ave{f_\al}(\bp;D) \equiv \lim_{r\ra\infty}\inv{\pi r^2}\int_{x_{\perp}'<r} d^2\mathbf{x_{\perp}'}
f_\al\left(D\hbz+\bx_{\perp}+\mathbf{x_{\perp}'},\bp,t\right).
\end{align}

Similarly, we assume that correlation functions of $f,B$ and $E$
may be defined, which are independent of $t$ and $\bx_\perp$, e.g.
\begin{align}\label{eq:Bcor}
&B_{ij}(\Dbx,\Dt,D)\equiv\cr&\lim_{r\ra\infty}\inv{\pi
r^2}\int_{x_{\perp}'<r} d^2\mathbf{x_{\perp}'}\cr
&B_i\left[D\hbz+\bx_{\perp}+\mathbf{x_{\perp}'},t\right]
B_j\left[D\hbz+\bx_{\perp}+\Dbx+\mathbf{x_{\perp}'},t+\Dt\right].\cr
\end{align}

The stationary shock structure is described by $\ave{f_\al}$ and by the (infinite) set of correlation functions. Note,
that the Maxwell-Vlasov equations may be converted to an (infinite) hierarchy of equations for the correlation functions.

It is useful for the analysis that follows to define the two-dimensional power spectrum of the magnetic field in a plane
parallel to the shock front. We define the 2D power spectrum at distance $D$ as
\begin{align}
B^{(2D)}_{ij}(\bk_{\perp},D) \equiv (2\pi)^{-1} \int
d^{2}\mathbf{x_{\perp}'}~B_{ij}(\mathbf{x_{\perp}'},D)e^{i\bk_{\perp}\cdot\mathbf{x_{\perp}'}},
\end{align}
where ${B_{ij}(\mathbf{x_{\perp}'},D)\equiv B_{ij}(\Dbx=\mathbf{x_{\perp}'},\Dt=0,D)}$. The plane averaged magnetic
energy density is related to the correlation function by
\begin{align}\label{eq:AveragedMagneticEnergy}
\ave{U_B}(D)&=(8\pi)^{-1}\sum_i (2\pi)^{-1} \int
d^2\mathbf{k_{\perp}}B^{(2D)}_{ii}(\bk_{\perp},D)\cr
&=(8\pi)^{-1}\sum_iB_{ii}(\Dbx=0,\Dt=0,D).
\end{align}

Note that the plane-averaged magnetic field vanishes. As a consequence of the planar symmetry, a non-zero averaged
magnetic field would be oriented along the $z$-axis. As its divergence vanishes, the magnitude of $\ave{\bB}$ would be
constant and thus equal to its far upstream value which is zero. The root mean square (rms) magnetic field,
\begin{align}\bar
B(D)=\left[\sum_iB_{ii}(\Dbx=0,\Dt=0,D)\right]^{1/2},\end{align}
serves as a measure of the characteristic magnetic field amplitude
at a distance $D$ from the shock (note, $\ave{U_B}=\bar
B^2/8\pi$).

\section{Self-similarity in the downstream}
\label{sec:self-sim}

As discussed in the introduction, afterglow observations suggest that the characteristic length scale $L$ for variations
in the magnetic field becomes much larger than $l_{sd}$, $L\gg l_{sd}$, at distances $D\gg l_{sd}$ downstream of the
shock. We assume here that in the limit $L/l_{sd}\rightarrow\infty$, $L$ becomes the only length scale relevant for the
evolution of the plasma, which implies self-similarity. There is no proof that self-similarity will be present whenever
the characteristic length scale diverges (or becomes infinitesimal). However, the self-similarity assumption is known to
be valid for many physical systems in which such divergence occurs [see, e.g., \citet{zel68} for self-similarity in
hydrodynamics, and \citet{Kadanoff67} for self-similarity in critical phenomena]. In order to clarify the reasoning
behind this assumption and its implications, we first briefly describe in \S~\ref{sec:examples} an example of a physical
system, where the presence of a diverging characteristic length scale leads to self-similar behavior. We then formulate
in \S~\ref{sec:DownStream} the self-similarity assumption for the downstream of collisionless shocks. The separation of
the plasma into two components is discussed in \S~\ref{sec:2comp}. The scaling relations are derived in
\S~\ref{ScalingRelations}, and some of their implications are discussed in \S~\ref{sec:implications}.

A note is in place here regarding non-relativistic collisionless
shocks. As mentioned in the introduction, self-similarity may be
expected in the non-relativistic shocks observed in young SNR's.
The analysis is similar for relativistic and non-relativistic
shocks, and in particular leads to the same scaling relations and
implications. The main difference is that in the non-relativistic
case there is a physically relevant length scale, which is
different than the skin depth $l_{sd}$: The Larmor radius of
mildly relativistic protons, $R_L\sim m_pc^2/eB\sim
R_{L,th}(\vs/c)^{-1}\sim (\vs/c)^{-1}l_{sd}$, where $R_{L,th}$ is
the characteristic Larmor radius of thermal protons. In other
words, there is a small parameter in the problem, $(\vs/c)$, which
is likely to have a physical implication. Hence, although in this
case too we expect self-similarity as $L/l_{sd}\rightarrow\infty$,
we expect the self-similar solution to provide a good
approximation for $L\gg(\vs/c)^{-1}l_{sd}$ rather than for $L\gg
l_{sd}$, and the scaling relations of the particle distribution
functions to be valid for momenta $p>m_p c$.

\subsection{An example of self-similarity}
\label{sec:examples}

Consider a macroscopic system in thermodynamic equilibrium, exhibiting a phase transition at some temperature $T_c$. Near
the critical temperature, $T\sim T_c$, the correlation length $L$ of fluctuations within the system diverges, $L(\delta
T)\rightarrow\infty$ as $\delta T\equiv T-T_c\rightarrow0$. It is commonly assumed that in this limit $L$ becomes the
only "relevant" length scale, and that the microscopic length scales, e.g. the inter-particle distance $d$, become
irrelevant.

To clarify this assumption, let us consider the correlation function $f(r)$ of some fluctuating quantity. $f$ is a
function of $\delta T$ and of the various parameters defining the thermodynamic system (e.g. $d$). Assuming $L$ diverges
monotonically with $\delta T$ we may replace the dependence on $\delta T$ with dependence on $L$, $f(r;L,d,c_i)$ where
$c_i$ represent the parameters defining the thermodynamic system. We may assume that these parameters contain no
parameter with the dimensions of length, since any such parameter $c_k$ may be replaced by a dimensionless parameter,
$c_k/d$. Using the dimensional parameters,
a constant $A_f$ with the same dimensions of $f$ may be constructed, and $f$ may be written as $f=A_f\tilde{f}$ where
$\tilde{f}$ is a dimensionless function.
Since $\tilde{f}$ is dimensionless, it must be a function of only dimensionless combinations of $\{r,L,d,c_i\}$. Thus,
$f=A_f\tilde{f}(r/L;d/L,\tilde{c}_i)$ where $\tilde{c}_i$ represent the dimensionless combinations of $c_i$. As $L/d$
diverges it is assumed that $d$ becomes "irrelevant," in the sense that $\tilde{f}$ approaches a limit
\begin{equation}\label{eq:slfsim}
    f(r;L,d,c_i)=A_f\tilde{f}\left(\frac{r}{L};\frac{d}{L},\tilde{c}_i\right)
    \xrightarrow[d/L\rightarrow 0]{}
    A_f\left(\frac{d}{L}\right)^{-s_f}\tilde{g}\left(\frac{r}{L};\tilde{c}_i\right).
\end{equation}
Note, that it is not assumed that, as $d/L\rightarrow 0$, $\tilde{f}$ approaches a limit independent of $d/L$, but rather
that in this limit it has a
scale-free (power-law)
dependence on $d/L$. In the former case, $f$ depends on the dimensional parameters $A_f$ and $L$ alone, and in the latter
on $A_fd^{-s_f}$ and $L$. In both cases, the only length scale that may be extracted from $f$ is $L$. Similar arguments
lead to the conclusion that $L$ has a power-law dependence on $\delta T$,
\begin{equation}\label{eq:slfsimL}
    \delta T(L,d,c_i)=T_c\tilde{\delta T}\left(\frac{d}{L},\tilde{c}_i\right)
    \xrightarrow[d/L\rightarrow 0]{}
    T_c\left(\frac{d}{L}\right)^{-s_T}\tilde{h}\left(\tilde{c}_i\right).
\end{equation}

In general, the self-similarity assumption may be stated as the assumption that all the properties of the system at some
$\delta T=\delta T_1$, corresponding to $L_1=L(\delta T_1)$, are identical to those of the system at $\delta T=\delta
T_2$, corresponding to $L_2=L(\delta T_2)$, up to a scaling
transformation.
That is, any function describing some properties of the system, e.g. $f[r,\delta T(L)]$, is identical at $L_1$ and $L_2$
up to a scaling of $r$ and $f$,
\begin{equation}\label{eq:slfsimf}
    f(r,L_2)=\left(\frac{L_2}{L_1}\right)^{s_f}f\left(\frac{r}{L_2/L_1},L_1\right)
\end{equation}
for some $s_f$. The function $f$ is termed self-similar, as it is similar to itself at different $L$ values. Comparing
Eqs.~(\ref{eq:slfsimf}) and~(\ref{eq:slfsim}) it is clear that the assumption that $d$ is "irrelevant" is equivalent to
the assumption that $f$ is self-similar.

It is clear from the above example that the assumption of self-similarity provides powerful constraints on the properties of the system, and that a complete characterization of its properties requires determination of the similarity exponents $\{s_f,s_T\}$.

\subsection{The self-similarity assumption}
\label{sec:DownStream}

Consider the plasma at a distance $D\gg l_{sd}$ downstream of the shock, where the field correlation length is assumed to
satisfy $L\gg l_{sd}$. Assuming $L/l_{sd}$ diverges with $D$, we expect the structure of the shock to become
self-similar. That is, we expect the averaged distribution function $\ave{f_\al}$, the rms magnetic field $\bar B$ and
the (infinite) set of correlation functions to be self-similar:
\begin{align}
\label{eq:slfB} \bar B(L)= \left(\frac{L}{L_0}\right)^{s_B} \bar
B(L_0),
\end{align}
\begin{align}
\label{eq:slfF} \ave{f_{\al}}(\bp,L)=
\left(\frac{L}{L_0}\right)^{s_f}
\ave{f_\al}\left[\frac{\bp}{(L/L_0)^{s_p}},L_0\right],
\end{align}
\begin{align}\label{eq:slfCorB}
B_{ij}(\Dbx,\Dt,L)=\left(\frac{L}{L_0}\right)^{2s_B}
B_{ij}\left[\frac{\Dbx}{L/L_0},\frac{\Dt}{(L/L_0)^{s_t}},L_0\right],
\end{align}
and so on,
where $L_0$ is a reference length scale.
Note, that we have replaced the dependence on $D$ with a dependence on $L$, assuming that $L$ diverges monotonically
with $D$.

The values of the similarity exponents are not determined by the requirement of self-similarity alone, and must be
derived based on additional arguments. In what follows, we derive constraints on the similarity exponents
$\{s_t,\,s_p,\,s_B...\}$ using the Maxwell-Vlasov equations. Since self-similarity applies to averaged quantities, e.g.
$\ave{f_\al}$ and $B_{ij}$, rather than to specific solutions, this derivation should be based on the equations for the
averaged quantities, which are obtained from the Maxwell-Vlasov equations. Such a derivation is, however, rather
cumbersome. Identical constraints on the similarity exponents may be derived in a simpler way by assuming that the
specific solutions form a scalable family in the sense that if $\{f_\al(\bx,\bp,t),\,\bB(\bx,t)\}$ is a solution of the
Maxwell-Vlasov equations, then the scaled functions
\begin{align}\label{SeSAssumptionBf}
&\bB'(\bx,t)=\xi^{s_B}\bB\left(\frac{\bx}{\xi},\frac{t}{\xi^{s_t}}\right),\cr
&{f_{\al}}'(\bx,\bp,t)=\xi^{s_f}f_{\al}\left(\frac{\bx}{\xi},\frac{\bp}{\xi^{s_p}},\frac{t}{\xi^{s_t}}\right),
\end{align}
also constitute a solution (at least approximately). It is straightforward to verify that the self-similarity
requirements, Eqs.~(\ref{eq:slfF})-(\ref{eq:slfCorB}), are automatically satisfied under this assumption, by substituting
$\xi=L/L_0$ in Eq.~(\ref{SeSAssumptionBf}) and recalling that the averages and correlation functions are identical for
all specific solutions. Under the assumption that the solutions are scalable, we can use the Maxwell-Vlasov equations
directly (instead of the equations for the correlation functions) in order to derive the constraints on the similarity
exponents. Before doing so, we address (in \S~\ref{sec:2comp}) the separation of the plasma into two components.

It is important to emphasize here, that although the constraints derived under the assumption that the solutions are
scalable (in the above sense) are identical to those derived from the self-similarity requirements,
Eqs.~(\ref{eq:slfF})-(\ref{eq:slfCorB}), it is not obvious that the self-similarity requirements indeed imply that the
specific solutions are scalable.

\subsection{Two plasma components}
\label{sec:2comp} As the plasma flows away from the shock, most of the particles remain "thermal," \footnote{The term
"thermal" is used for convenience to describe particles with kinetic energy $\sim (\gamma_s-1)m_pc^2$. In what follows,
it is not claimed or assumed that the energy distribution of these particles is thermal i.e. a (relativistic) Maxwellian
distribution.} i.e. most protons carry a momentum $\sim p_{th}$ (and most electrons carry a momentum $(m_e/m_p)^\nu
p_{th}$ with $\nu\ge0$). This implies the existence of a length scale $R_{L,th}(L)=p_{th} c/e\bar B\propto L^{-s_B}$,
$R_{L,th}/L\propto L^{-(1+s_B)}$. Obviously, in order for $L$ to be the only relevant scale, we must have $s_B>-1$.
Combined with the requirement that the field energy density does not diverge we have
\begin{equation}\label{eq:sB}
    -1<s_B\le 0.
\end{equation}

Our assumption that $L$ is the only relevant length scale implies that $R_{L,th}$ is irrelevant. However, $R_{L,th}$ is
obviously relevant for the description of the microscopic motion of individual "thermal" particles. This apparent
contradiction may be resolved only if the microscopic motion of "thermal" particles, i.e. of particles with momenta $p$
such that $pc/e\bar B\ll L$, is unimportant. These particles should therefore be described by effective equations, where
the microscopic particle motion is unimportant. We leave the discussion of the "thermal" particles, to which we refer
henceforth as the "fluid" particles, to \S~\ref{sec:fluid}, and note here only several points which are important for the
discussion that follows.

First, the scaling of $f_\alpha$ derived in the previous section does not apply to all particle momenta, but rather to
the large $p$,
 $p\gg p_{th}(\vs/c)^{-1}$,
 behavior of $f_\al$.
 In particular, the Vlasov equation in this region, where
 $\bbv(\bp)\approx c\hbp$, can be written as,
\begin{equation}\label{eq:VlasovRel}
\pr_tf_\al+c\hbp\cdot\nabla f_\al+q_\al\left(\bE+\inv{c}\bbv\X\bB\right)\cdot\nabla_{\bp}f_{\al}=0.\end{equation}
 Second, the total electric current is the sum of the electric currents carried by the high energy
particles, $\bj_h$, and by the "fluid," $\bj_F$, $\bj=\bj_h+\bj_F$. Hereafter, $h$ and $F$ subscripts refer to the high
energy and to the fluid particles, respectively. The electric current $\bj_h$ is given by
\begin{align}\label{eq:accellcur}
\bj_h=\sum_\al q_\al\int_{p>\xi_p e\bar BL/c} d^3\bp~c~\hbp
f_\al(\bp).
\end{align}
Here, $\xi_p(L)$ is a dimensionless function that determines the threshold momentum between accelerated and fluid
components. It must satisfy $\xi_p>(\vs/c)^{-1}p_{th}c/e\bar B L=(\vs/c)^{-1}R_{L,th}/L$ in order to ensure that the
accelerated component is not affected by the thermal scale. It must also tend to zero as $L$ diverges, as self-similarity
is expected for all scales $\gg (\vs/c)^{-1}R_{L,th}$.

Solutions of the type we have arrived at, where the self-similar solution describes the evolution in some part of
($\bx,\bp,t$) phase space while the other part is described by a different solution, are usually called "second type
solutions" \citep[e.g.][]{zel68,wax93}. In second type solutions, the similarity exponents can not be determined by
global conservation laws. For example, one may have argued that a relation between $s_f$ and $s_p$ may be derived by
requiring the integral over momenta of $\ave{f}$, given by Eq.~(\ref{eq:slfF}), to be equal to the (downstream) particle
density,
\begin{equation}\label{eq:numberInt}
n_\al=\int d^3p \ave{f_\al}(\bp,L)=\left(\frac{L}{L_0}\right)^{3s_p+s_f}\int d^3y\ave{f_\al}(\textbf{y},L_0),
\end{equation}
which would imply $3s_p+s_f=0$. Such a constraint can not be derived, however, since the scaling relation,
Eq.~(\ref{eq:slfF}), used for obtaining the second equality of Eq.~(\ref{eq:numberInt}) does not hold for small values of
$y$. In fact, as we show below, the functional form of $f$ derived by the self-similarity arguments is such that the
integral on the rhs of Eq.~(\ref{eq:numberInt}) diverges at small $y$. The number of particles described by the
self-similar solution does not diverge, however, since the integration extends only down to
$y=\xi_p(L)(L/L_0)^{1-s_p}e\bar BL_0/c$ (note, that this constrains the dependence of $\xi_p$ on $L$). This is analogous
to the divergence of energy in second-type self-similar solutions of hydrodynamic flows \citep[e.g.][]{wax93}.

The particles with $p\lesssim\xi_p(L) e\bar BL/c$ are described by a solution different than the self-similar solution
describing the higher momenta, "accelerated," particles. As in any second type self-similar solution, it is necessary
that the non self-similar part of the solution does not affect the self-similar part
\citep[this requirement often determines the similarity exponents of second type solutions, e.g. ][]{zel68,wax93}. In our case, the particles at $p\lesssim\xi_p(L) e\bar BL/c$ may
affect the higher momentum particles only through their contribution to the electric current. This implies that the
current contributed by $p\lesssim\xi_p(L) e\bar BL/c$ particles must either be negligible or scale with $L$ in a similar
way as the current of the higher momenta particles does. For the fluid current, this requirement implies that either
$j_F\ll j_h$ or $j_F(L)/j_h(L)\propto L^0$. In addition, this requirement implies that the integral on the rhs of
Eq.~(\ref{eq:accellcur}) converges (as otherwise the current would be dominated by $p\lesssim\xi_p(L) e\bar BL/c$
particles).

The following point should be emphasized here. We have implicitly assumed above that the high momenta, accelerated
particles are dynamically important in the sense that their electric current makes a considerable contribution to the
total electric current, $\bj_h\sim\bj$. However, self-similar solutions where the accelerated particles do not contribute
to the current, $j_h\ll j$, are possible in principle. In this case, the growth of the magnetic field fluctuation
correlation length should be driven by the fluid, and the accelerated particles can be treated as "test-particles," which
do not affect the flow. As mentioned in the introduction, such a picture is unlikely, due to the fact that the
accelerated particles carry a significant fraction of the energy and due to the evidence that they play a role in
generating magnetic fields in the upstream plasma.

\subsection{Scaling relations}
\label{ScalingRelations}

Let us consider first the scaling of $L$ with $D$. Since the characteristic length scale for changes in $L$ is $L$, we
must have $L\propto D$: Requiring the scale for changes in $L$ to be proportional to $L$ may be written as $d(\log
L)/dD\propto1/L$ which implies $L\propto D$. It should be noted that this assumption is not equivalent to assuming that
the only length scale relevant for the evolution is $D$. Under such an assumption, $d(\log L)/dD=s/D$ implying $L=D^s$,
which allows the possibility $L/D\ra0$ as $D\ra\infty$.

Next, we consider the scaling of time, i.e. the value of $s_t$. Substituting the scaled solutions,
Eqs.~(\ref{SeSAssumptionBf}), into the Valsov equation, Eq.~(\ref{eq:VlasovRel}), one finds that the various terms in the
equation scale differently with $\xi$. In order for the scaling of the temporal and the spatial derivative terms to be
similar, one must require $s_t=1$. However, one can not conclude from this that $s_t=1$ is the only value allowed, since
it is possible that one of the first two terms, $\pr_t f$ or $\bbv\cdot\nabla f$, becomes negligible as
$L\rightarrow\infty$, and may be neglected altogether in the Vlasov equation. Nevertheless, we conclude that $s_t=1$ is
required based on the following arguments. The case $s_t<1$, implying that $cT/L\ra0$ as $L\ra\infty$ where $T$ is the
characteristic time for variations in the physical quantities, is ruled out since it would imply that at large distances
from the shock the electric field is much larger than the magnetic field ($E/B\ra\infty$ as $L\ra\infty$), which is
inconsistent with the synchrotron model of afterglow emission. Assuming $s_t>1$, $cT/L\ra\infty$ as $L\ra\infty$, would
imply that at large distances from the shock the magnetic fields are approximately static (velocities of the field lines
of the order $L/T$) in the shock frame. Once the condition $L/T\ll \vs$ would be reached, the thermal protons would
gyrate around the static field lines (the gyration radius, $R_{L,th}$, is much smaller than $L$, and thus also the
gyration time, $T_{L,th}\sim R_{L,th}/\vs,$
 is much shorter than $T$), in contradiction with the fact that the bulk downstream fluid moves
with velocity $\vd\lesssim \vs$ (Note that $\gamma_d=(1-\vd^2/c^2)^{-1/2}\sim1$). We conclude that we must have $s_t=1$.

The scaling of the momentum, i.e. the value of $s_p$, is determined by comparing the momentum derivative term with the
spatial (or temporal) derivative term in the Vlasov equation. Substituting the scaled solutions,
Eqs.~(\ref{SeSAssumptionBf}), into the Vlasov equation, Eq.~(\ref{eq:VlasovRel}), and requiring all terms to scale
similarly implies $s_p=1+s_B$. One implication of this scaling is that the Larmor radius scales as $L$ (in fact, the
entire trajectory of each particle scales with $L$).

Finally, we derive the relation between $s_f$ and $s_B$. The current provided by the accelerated particles is
\begin{eqnarray}
\label{eq:slf_current}
\bj'_h(\bx,t)&=&\sum_\al q_\al\int d^3\bp~c~\hbp \xi^{s_f}f_{\al}\left(\frac{\bx}{\xi},\frac{\bp}{\xi^{s_B+1}},\frac{t}{\xi}\right)\nonumber\\
&=&\xi^{s_f+3(s_B+1)}\sum_\al q_\al\int d^3\bp~c~\hbp f_{\al}\left(\frac{\bx}{\xi},\bp,\frac{t}{\xi}\right)\nonumber\\
&=&\xi^{s_f+3(s_B+1)}\bj_h\left(\frac{\bx}{\xi},\frac{t}{\xi}\right).
\end{eqnarray}
We have not shown here explicitly the lower limit of the integration over $p$, since, as discussed in \S~\ref{sec:2comp},
the current integral must converge at small $p$. Substituting the scaled current, and the scaled magnetic field, into
Maxwell's equation, $\nabla\X\bB=4\pi c^{-1}\bj+c^{-1}\pr_t\bE$, we find that the $\nabla\X\bB$ and $\bj$ terms scale
similarly with $\xi$ provided that $s_f+3(s_B+1)=s_B-1$.

To summarize, in order for the scaled solutions, Eqs.~(\ref{SeSAssumptionBf}), to satisfy the Vlasov-Maxwell equations,
it is necessary (and sufficient) for the similarity indices to satisfy
\begin{align}
\label{eq:indices}
s_t=1,\quad s_p=s_B+1,\quad s_f=-4-2s_B.
\end{align}
In addition, we have shown that $-1<s_B\le0$ and that
\begin{align}
\label{eq:LD}
L\propto D.
\end{align}

The relation between $s_p$ and $s_B$ implies that the Larmor radius of the particles scales as $L$. The relation between
$s_f$ and $s_B$ implies that the energy density of accelerated particles in any momentum interval,
$U_{h}=\sum_{\alpha}\int d^{3}\bp cpf_{\alpha}$, scales as the magnetic field energy density (when the momentum interval
scales appropriately). Combining the above results, we find that $\ave{U_h} \propto \ave{U_B} \propto L^{2s_B} \propto
D^{2s_B}$.

\begin{figure}[h]
\centerline{
\includegraphics[angle=-90,width=3.5in]{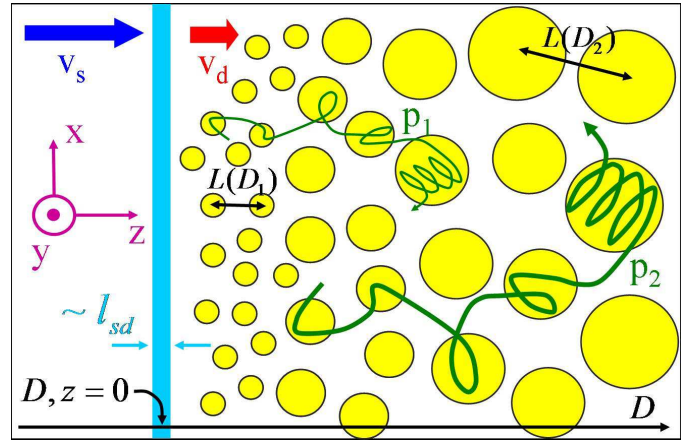}
} \caption{\label{fig:Illustration} Schematic illustration of the model. The shaded vertical band at $D=z=0$ represents
the shock. Shaded Circles designate regions where the magnetic field amplitude $B$ exceeds some constant threshold. The
characteristic scale $L$ for variations in $B$ increases with distance $D$ from the shock. Curved lines illustrate the
trajectories of two high-energy particles of different momenta, $p_1<p_2$, becoming magnetized ($R_L<L$) at two different
scales.}
\end{figure}

An illustration of the plasma configuration downstream of a hypothetical self-similar collisionless shock appears in
Figure~(\ref{fig:Illustration}).

\subsection{Additional implications}
\label{sec:implications}
The scaling relations derived above may be used to constrain the momentum dependence of the particle distribution
functions. Using $L_0=L(p_0/p)^{1/s_p}$ in Eq.~(\ref{eq:slfF}) we obtain
\begin{eqnarray}
\label{eq:highp}
\ave{f_{\al}}(\bp,L)&= \left(\frac{p}{p_0}\right)^{s_f/s_p}
\ave{f_\al}\left[\hbp p_0,L\left(\frac{p_0}{p}\right)^{1/s_p}\right]\nonumber\\
&\xrightarrow[p/p_0\rightarrow\infty]{}\ave{f_\al}\left(\hbp
p_0,0\right)\left(\frac{p}{p_0}\right)^{s_f/s_p}.
\end{eqnarray}
This implies that $\ave{f_{\al}}(\bp,L)$ approaches a power-law distribution at large $p$, $\ave{f_{\al}}(\bp,L)\propto
p^{s_f/s_p}$, provided that $\ave{f_{\al}}(\bp,L)$ approaches a finite limit for some $p$ as $L\rightarrow0$. The
condition that $\ave{f_{\al}}(\bp,L)\rightarrow{\rm Const}\neq0$ as $L\rightarrow0$ is equivalent to the requirement that
accelerated particles reach the shock front. This is physically reasonable since cosmic rays are expected to reach the
upstream. Moreover, it is a necessary condition for any acceleration process that involves the shock front (such as
diffusive Fermi acceleration).

A similar constraint may be obtained by taking the limit $p/p_0\rightarrow0$, which yields
\begin{eqnarray}
\label{eq:lowp}
\ave{f_{\al}}(\bp,L)&= \left(\frac{p}{p_0}\right)^{s_f/s_p}
\ave{f_\al}\left[\hbp p_0,L\left(\frac{p_0}{p}\right)^{1/s_p}\right]\nonumber\\
&\xrightarrow[p/p_0\rightarrow0]{}\ave{f_\al}\left(\hbp
p_0,\infty\right)\left(\frac{p}{p_0}\right)^{s_f/s_p}.
\end{eqnarray}
This implies that $\ave{f_{\al}}(\bp,L)$ approaches a power-law distribution at small $p$, $\ave{f_{\al}}(\bp,L)\propto
p^{s_f/s_p}$, provided that $\ave{f_{\al}}(\bp,L)$ approaches a finite limit for some $p$ as $L\rightarrow\infty$. The
condition that $\ave{f_{\al}}(\bp,L)\rightarrow{\rm Const}\neq0$ as $L\rightarrow\infty$ is equivalent to the requirement
that accelerated particles reach downstream infinity. This is physically reasonable since cosmic rays are expected to be
advected with the flow to the downstream (e.g. in diffusive Fermi acceleration). Moreover, the presence of high energy
electrons in the far downstream ($\sim 10^{10}l_{sd}$) is required to account for afterglow observations. Thus, we
conclude that the self-similar distribution function approaches a power-law dependence on $p$ at both high and low
momenta,
\begin{equation}
\label{eq:dndp} \frac{dn_{\al}}{dp}\propto p^2\int d^2\hbp f_{\al}(p\hbp)\propto p^{-2/(s_B+1)}.
\end{equation}

Let us consider next the magnetic field correlation function. Using $L/L_0=\Delta x/\Delta x_0$ in Eq.~(\ref{eq:slfCorB})
we obtain
\begin{eqnarray}
\label{eq:CorrBscaling}
B_{ij}(\Dbx,\Dt,L)&=\left(\frac{\Delta x}{\Delta x_0}\right)^{2s_B}
B_{ij}\left[\Delta x_0\hat{x},\frac{\Dt}{\Delta x/\Delta x_0},\frac{\Delta x_0}{\Delta x}L\right],\nonumber\\
&\xrightarrow[\Delta x/\Delta
x_0\rightarrow\infty]{}\left(\frac{\Delta x}{\Delta
x_0}\right)^{2s_B}B_{ij}\left(\Delta x_0\hat{x},0,0\right),
\end{eqnarray}
where $\hat{x}$ is a unit vector in the direction of $\Dbx$. This implies that $B_{ij}(\Dbx,\Dt,L)$ approaches a
power-law behavior at large $\Delta x$, $B_{ij}(\Dbx,\Dt,L)\propto \Delta x^{2s_B}$, provided that $B_{ij}(\Dbx,\Dt,L)$
approaches a finite limit for some $\Delta x_0$ as $L,\Dt\ra0$. This is expected to be the case under the assumption that
the accelerated particles reach the shock front, since in this case the number of particles with Larmor radius greater
than (some given) $\Dx$ approaches a constant as $L\ra 0$, and such particles are expected to generate a correlation
between the magnetic fields at points separated by $\Dbx$.

Eq.~(\ref{eq:CorrBscaling}) implies that the magnetic power spectrum has a component which scales as $k^{-2-2s_{B}}$.
Assuming that the result $B_{ij}(\Dbx,0,L)\approx\left(\Delta x/\Delta x_0\right)^{2s_B}B_{ij}\left(\Delta
x_0\hat{x},0,0\right)$ holds for $\Dx>\eta \Delta x_0$,
where $\eta$ is some dimensionless constant,
 the power spectrum may be written as (see \S~\ref{sec:stationary})
\begin{align}
&2\pi B^{(2D)}_{ij}(\bk_{\perp},L)\approx
\int_{x_{\perp}'<\eta\Delta x_0} d^{2}\mathbf{x_{\perp}'}~
B_{ij}(\mathbf{x_{\perp}'},L)e^{-i\mathbf{k_{\perp}\cdot
\mathbf{x_{\perp}'}}}\cr &+(\Delta x_0)^{-2s_B}\int_{\eta\Delta
x_0}^{\infty}dx_{\perp}'~ x_{\perp}'^{1+2s_{B}}\int_0^{2\pi} d\te~
B_{ij}(\Delta x_0\mathbf{\hat
x'_{\perp}},0,0)e^{-i\mathbf{k_{\perp}\cdot\mathbf{x_{\perp}'}
 }}\cr
 &= f_{1}\left(\mathbf{k_{\perp}},\eta\Delta x_0,L\right)\cr
 &+(\Delta
x_0)^{-2s_B}k_{\perp}^{-2-2s_{B}}\int_{k_{\perp}\eta\Delta
x_0}^{\infty}du~ u^{1+2s_{B}}\int_0^{2\pi} d\te~B_{ij}(\Delta
x_0\mathbf{\hat{u}},0,0)e^{-i\mathbf{\hat{k}_{\perp}\cdot
 u}},\cr
\end{align}
where, $\mathbf{u}\equiv k_{\perp}~\mathbf{x_{\perp}'}$. In the long wavelength limit,
$0<k_{\perp}\ll\left(\eta\Delta x_0\right)^{-1}$, the exponent in $f_{1}$ is approximately constant, such
that $f_{1}$ is nearly independent of $\mathbf{k_{\perp}}$. In addition, the lower limit in the second integral,
$k_{\perp}\eta\Delta x_0$, may be taken to zero (recall that $s_{B}>-1)$. Thus, $2\pi B^{(2D)}_{ij}(\bk_{\perp},L)\approx
f_{1}(\eta\Delta x_0,L)+k_{\perp}^{-2-2s_{B}}f_2(\mathbf{\hat k_{\perp}},\Delta x_0)$ where we assumed that the integral
in the second term converges. As $s_B>-1$, for $k_{\perp}\ra 0$ we can neglect $f_{1}$ so that
\begin{align}\label{eq:magnetic_power_spectrum}
B^{(2D)}_{ij}(\bk_{\perp},L)\propto
k_{\perp}^{-2-2s_{B}}.\end{align}

The scaling of various other correlation functions, such as $\ave{f_{\al}f_{\beta}}$, may be deduced in the method shown
above. For example, the arguments leading to the conclusion $B_{ij}(\Delta\bx,\Delta t,L)\propto \Delta x^{2 s_B}$ for
$\Dx\ra \infty$ also imply that $B_{ij}(\Delta\bx,\Delta t,L)\propto \Delta t^{2 s_B}$ for $\Delta t\ra\infty.$

\section{The "fluid" component, downstream}
\label{sec:fluid}

As explained in \S\ref{sec:2comp}, the assumption that $L$ is the only relevant scale implies that the microscopic motion
of the thermal particles (referred to as the "fluid" particles) is unimportant for the global solution. One option is
that the current carried by the fluid is negligible, $j_F\ll j_h$, in which case these particles do not affect the EM
fields. The second option is that the current carried by this population is considerable $j_F\sim j_h\sim j$ (i.e. that
it scales as the electric current carried by the accelerated component), and that the thermal fluid is described by
effective equations independent of the microscopic scales $l_{sd}$ and $R_{L,th}=p_{th}c/eB$. The first option is
unlikely since, as we show in \S\ref{sec:InfiniteCon}, the fluid is expected to be highly conducting and thus probably
carries non-negligible currents. In this section we study the implications of the second option, $j_F\sim j_h$. We assume
that the appropriate effective degrees of freedom are those of a single fluid, since the skin depth $l_{sd}$ and thermal
Larmor radius $R_{L,th}$ have an important role already at the level of two fluid equations.

In order to obtain equations which are independent of the downstream bulk velocity of the fluid, which can not scale with
$L$, we consider the flow in the downstream frame (where the bulk velocity vanishes). Note, that since $s_t=1$, the self
similarity relations, e.g. Eqs. \eqref{SeSAssumptionBf}, hold both in the shock frame and in the downstream frame (and in
any other frame, as long as we set $t=0$ as the time when the shock was at $z=0$). As the current carried by the fluid is
non-negligible, $\bj_F\sim \bj$, it must have a self similar structure. We thus assume that all the dynamically important
fluid quantities have a self similar structure. In particular, the fluid velocity field $\bu$ (in the downstream frame)
scales as
\begin{align}\label{eq:slfu1}
\bu'(\bx,t)=\xi^{s_u}\bu(\frac{\bx}{\xi},\frac{t}{\xi}).
\end{align}

In \S\ref{sec:InfiniteCon} we show, that for $L\gg R_{Lth}$ the fluid can be considered as infinitely conducting, i.e.
\begin{align}\label{eq:InfiniteCon}
\bE=-\inv{c}\bu\X\bB.
\end{align}
This, together with Maxwell's equation $\nabla\X\bE=-c^{-1}\pr_t\bB$, implies that
$L/T\sim u$ and thus
$s_u=1-s_t=0$. In other words, the velocity fluctuations scale trivially. In order to relate this to the magnetic field
we consider the force density applied to the fluid, $\sim \gamma_s^2n_1m_pu/T$ where $T$ is the characteristic time scale
for variations. The force has a contribution from the pressure, $\sim P/L$ where $P$ is the fluctuating part of the
pressure, and from the EM fields,
$\sim c^{-1}j_fB \sim U_B/L$ where $U_B$ is the magnetic field energy density. If $U_B$ is not negligible compared to $P$,
requiring that the two force densities, $\gamma_s^2n_1m_pu/T$ and
$c^{-1}j_fB
\sim U_B/L$,
scale similarly we find, using $u\propto L^0$ and $T\propto L^1$, that $s_B=0$. If $U_B$ is negligible, $P\gg U_B$, the
force density is dominated by $P/L$ and $s_B$ can not be constrained by considering the force density scaling. However,
$P\gg U_B$ is unlikely. Our basic assumption that the accelerated component plays an important role in the solution
suggests that the fluctuations in the fluid are driven by it. As the interaction between the accelerated component and
the fluid is mediated through the EM fields this implies that they have a considerable contribution to the force density
applied to the fluid. Moreover, the observational evidence that $\ep_B$ is not very small,
$\ep_B\sim10^{-2}-10^{-1}$,
does not leave much room for $P$ (which is smaller than $U_B/\ep_B$) to be much larger than $U_B$. We conclude therefore
that
\begin{equation}\label{eq:su}
    s_u=s_B=0,
\end{equation} which, using Eq.~\eqref{eq:indices}, implies
\begin{align}
\label{eq:indices2} \quad s_p=1,\quad s_f=-4.
\end{align}
Eq. \eqref{eq:su} and \eqref{eq:indices2} imply that $U_u\propto U_B\propto U_h$, where $U_u$, $U_B$ and $U_h$ are the
energy densities of the fluid fluctuations, magnetic fields and accelerated particles, respectively. Implications
regarding the correlation functions of fluid quantities such as $\bu$ can be drawn in analogy to
\S~\ref{sec:implications}.

In \S\ref{sec:fluidEqs} below we study the scaling properties of the fluid equations of motion, under the simplifying
assumptions of an ideal fluid and that corrections of order $\ep_B$ can be neglected. A closed set of equations is
derived for the fluid component, and it is demonstrated that self-similar (scalable) solutions of the equations exist
only for $s_u=s_B=0$. In \S\ref{sec:InfiniteCon} we argue that under the assumption $R_{L,th}\ll L$, the fluid can be
treated as infinitely conducting (as we show, the exact requirement is $R_{L,th}/L\ll\{\vs T/L,~L/\vs T\}$).

One consequence of the scaling indices derived above is a divergence of the energy associated with the self-similar
components. A flat power-law energy-spectrum of accelerated particles, $dn/dE\propto E^{-2}$ [Eq. \eqref{eq:dndp} with
$s_B=0$], which is a natural consequence of the model and agrees with observations and with linear DSA theory, implies
that the energy density diverges at large momenta. Similarly, a flat magnetic power-spectrum of the form
$B^{(2D)}_{ij}(\bk_{\perp},L)\propto k_{\perp}^{-2}$ [Eq. \eqref{eq:magnetic_power_spectrum} with $s_B=0$] implies that
the magnetic energy density diverges at large wavelengths (the same applies to the kinetic energy associated with the
velocity fluctuations). Possible remedies for the divergence problem are outlined in \S~\ref{sec:divEng}, but resolving
the problem is beyond the scope of this paper.

\vfill
\subsection{Fluid equations}
\label{sec:fluidEqs}

For simplicity, we focus below on relativistic shocks. We comment at the end of this subsection on the application of the analysis to non-relativistic shocks. As mentioned above, we consider the flow in the downstream frame.

The total electric current is $\bj=\bj_F+\bj_h$. The fluid current may be written as
\begin{align}\label{eq:FluidCurrent}
\bj_F=\frac{c}{4\pi}\nabla\X\bB-\bj_h-\inv{4\pi}\pr_t\bE.
\end{align}
In order to guarantee conservation of charge we must include
\begin{align}\label{eq:charge}
\nabla\cdot\bE=4\pi(\rho_h+\rho_F)
\end{align}
as an independent equation, where $\rho_h$ is the electric charge density carried by the accelerated component,
\begin{align}\label{eq:rhoh}
\rho_h=\sum_\al q_\al\int_{p>\xi_p e\bar BL/c} d^3\bp f_\al,
\end{align}
and $\rho_F$ is the electric charge density carried by the fluid. Conservation of fluid energy and momentum is described
by
\begin{align}\label{eq:EnergyMomentum}
\pr_\nu T_F^{\mu\nu}=F^{\mu\nu}j_{F\nu},
\end{align}
where $T_F^{\mu\nu}$ is the fluid energy momentum tensor, $j_{F\nu}$ is the fluid four current, and $F^{\mu\nu}$ is the
electromagnetic tensor. The fluid velocity $\bu$ is defined so that $T^{0i}=0$ in a local frame moving with velocity
$\bu$. These equations can be closed with the addition of an "equation of state," relating the different components of
$T^{\mu\nu}$ in the fluid rest frame.

As the energy density of the fluid must scale as $L^0$, the above equations are consistent only with a scaling in which
the EM fields and $T^{\mu\nu}_F$ both scale as $L^0$, implying $s_u=s_B=0$. However, since the equations may contain
terms that can be neglected, and the scaling laws are likely to be relevant to fluctuations in $T^{\mu\nu}$ (imposed on a
constant background) rather than to its average value, different scaling exponents may also be allowed. In order to
examine this in more detail, we make the simplifying assumption of an ideal fluid, i.e. $T^{ij}\propto\de^{ij}$ in the
fluid rest frame. This is the case if the distribution of the fluid particles is approximately isotropic in the fluid's
local rest frame, which is reasonable as their Larmor radius is much smaller than the length scale for variations in the
magnetic field.
 Under this assumption, equation \eqref{eq:EnergyMomentum} can be written as
\begin{align}\label{eq:FullMHDEnergy}
\pr_t\left[\gamma^2(w_0+w)\right]+\nabla\cdot\left[\gamma^2(w_0+w)\bu\right]=\bE\cdot\bj_F+\pr_tP
\end{align}
and
\begin{eqnarray}\label{eq:FullMHDMomentum}
\nonumber\frac{\gamma^2(w_0+w)}{c^2}(\pr_t+\bu\cdot\nabla)\bu &+&
\frac{\bE\cdot\bj_F+\pr_tP}{c^2}\bu \\
& = & -\nabla P+\rho_F\bE+\inv{c}\bj_F\X\bB,
\end{eqnarray}
where $P_0+P$ and $w_0+w$ are the pressure and rest frame enthalpy per unit volume, respectively, assumed to consist of a
constant part, $P_0\sim w_0\sim \gamma_s^2n_1m_pc^2$, and a fluctuating part, $P,\,w$. Here,
$\gamma\equiv(1-u^2/c^2)^{-1/2}$ is the Lorentz factor associated with the fluid's velocity fluctuations. The fluid
equations must be simplified by identifying terms that are negligible, in order to determine the correct scaling laws. As
explained above, we assume that the magnetic force density, $c^{-1}|\bj_F\X\bB|\sim L^{-1} B^2\sim
L^{-1}\ep_B\gamma_s^2n_1m_pc^2$, makes a considerable contribution to the momentum balance [Eq.
\eqref{eq:FullMHDMomentum}]. This implies $P\lesssim\ep_B\gamma_s^2n_1m_pc^2\sim\ep_B P_0$. We assume that $w/w_0\sim
P/P_0$, which holds for example for any equation of state where the enthalpy is a function of the pressure alone, in
particular this is true for a relativistic equation of state, $w_0+w=4(P_0+P)$. We find
$w\lesssim\ep_B\gamma_s^2n_1m_pc^2\sim \ep_B w_0$. Comparing the first and the last terms in
Eq.~\eqref{eq:FullMHDMomentum}, combined with the above result, $L/T\sim u$, we find that $\gamma^2u^2/c^2\sim\ep_B$ and
thus $\gamma\approx 1$ and the fluid fluctuations are non-relativistic. We thus find that the thermal energy $P$ and the
kinetic energy $U_u\sim w_0u^2$ of the fluid fluctuations are related to the magnetic energy by $P,U_u\lesssim U_B$. Note
that $\bE\cdot\bj_F\sim uB^2/L~$ [since $Ej_F\sim (u/cB)(cB/L)$]. Hence, equations \eqref{eq:FullMHDEnergy} and
\eqref{eq:FullMHDMomentum} can be approximately written as
\begin{align}\label{eq:Velocity1}
&w_0\nabla\cdot\bu=0
\end{align}
and
\begin{align}\label{eq:FluidMomentumEquation}
&\frac{w_0}{c^2}(\pr_t+\bu\cdot\nabla)\bu=-\nabla
P+\rho_F\bE+\inv{c}\bj_F\X\bB.
\end{align}
In this case, we may eliminate the pressure from the equations by taking the curl of Eq.
\eqref{eq:FluidMomentumEquation}, yielding
\begin{align}\label{eq:velocityequaion2}
&\frac{w_0}{c^2}\left\{\pr_t\nabla\X\bu-\nabla\X\left[\bu\X(\nabla\X\bu)\right]\right\}=
\nabla\X(\rho_F\bE+\inv{c}\bj_F\X\bB).
\end{align}
Equations \eqref{eq:Maxwell}, \eqref{eq:VlasovRel}-\eqref{eq:accellcur}, \eqref{eq:InfiniteCon}, \eqref{eq:Velocity1} and
\eqref{eq:velocityequaion2}, constitute a closed set of equations for the unknowns $\bu,f_\al$ (for momentum $p>\xi_p
e\bar BL/c$) and $\bB$. As can be readily seen, these equations imply (and are consistent with) a scaling $s_B=s_u=0$.

The above analysis is also applicable to non-relativistic shocks.
A few comments should, however, be added. In the non-relativistic
case, the electric field is much weaker than the magnetic field,
$E\sim (u/c)B\ll (\vs/c)B\ll B$. In addition, it is expected that
$T\sim L/u\ll L/c$. Thus, the terms $\pr_t f_\al$ (for
relativistic momentum) and $q_\al\bE$ in Vlasov's equation
Eq.~\eqref{eq:VlasovRel}, and the term $c^{-1}\pr_t\bE$ in
Maxwell's equation ${\nabla\X\bB=4\pi c^{-1}\bj+c^{-1}\pr_t\bE}$,
may be neglected. Since the particles of the thermal component are
non-relativistic in this case, $w_0/c^2$ may be replaced in
equations~\eqref{eq:Velocity1} and~\eqref{eq:velocityequaion2} by
its non-relativistic limit, $\rho_{Fm_0}$ (mass density). The
assumption $w/w_0\lesssim P/P_0$, which for non-relativistic
particles reads $\rho_{Fm}/\rho_{Fm0}\lesssim P/P_0$, is somewhat
less justified but still reasonable. As mentioned above, these
differences do not modify the scaling relations.

\subsection{Infinite fluid conductivity}
\label{sec:InfiniteCon}

Here we argue that the assumption $R_{L,th}\ll L$ implies that the fluid can be treated as infinitely conducting (as we
show below the exact requirement is $R_{L,th}/L\ll\{\vs T/L,~L/\vs T\}$).

The fluid velocity is approximately equal to the the velocity of the proton component of the fluid.
We now demonstrate
that the electric and the magnetic forces acting on the proton component approximately cancel each other and thus lead to
Eq. \eqref{eq:InfiniteCon}. Consider the proton fluid in a box of size $<L$ during a time scale $<T$, in which the
magnetic and electric fields are approximately constant in space and time. The proton momentum inside the box is
$\sim\gamma_s m_pnuL^3$. The time derivative of the momentum in the box originates from the electric force, $f_E\sim
enEL^3\sim en(L/cT)BL^3$, the magnetic force, $f_B\sim enu/cBL^3$, and the flow of momentum into the box, $f_P\lesssim
L^2n(\gamma_s m_p\vs)\vs$. The ratio of the momentum flow to the electric force is of the order of
\begin{align}
\frac{f_P}{f_E}\lesssim \frac{\gamma_s
m_p\vs^2}{eBL}\frac{cT}{L}\sim\frac{R_{L,th}}{L}\frac{\vs T}{L}\ll1.
\end{align}
Hence, the only force that can balance $f_E$ is $f_B$. The forces are balanced only if $\bE=-c^{-1}\bu\X\bB$. If the
forces are not balanced, the proton fluid is accelerated. The time it takes the velocity $u$ to reach the scale $L/T$
(for which $\bE\sim-c^{-1}\bu\X\bB$), is very short, of the order of $\gamma_sm_pn(L/T)L^3/enEL^3\sim(R_{L,th}/\vs T)T\ll
T$. Therefore, we may assume that the forces are approximately balanced at any given time, and the fluid is nearly infinitely conducting.

\subsection{Diverging Energy}
\label{sec:divEng}

As explained in the beginning of this section, the vanishing of the magnetic scaling index, $s_B$, implies logarithmic
energy divergences. The divergence may be prevented if the scaling relations given in Eqs. \eqref{SeSAssumptionBf} and
\eqref{eq:slfu1} are approximate, rather than accurate. For example, high-order terms, $\propto1/L,\,1/L^2...$, that were
neglected in our analysis may introduce logarithmic corrections to the scaling relations. Alternatively, physical
processes that were not included in the analysis, such as cooling, may limit the parameter range over which
self-similarity is applicable, e.g. self-similarity may hold up to some cutoff momentum $p_{max}$ or cutoff distance
$D_{max}$.

Modeling GRB external shocks \citep{Waxman95} and SNR shocks \citep{Zhang93} implies that in both systems the maximal
energy of accelerated ions is limited by the age (size) of the system, whereas the maximal energy of the accelerated
electrons is limited by energy losses. If the accelerated electrons play an important role in the evolution,
self-similarity may break-down at some scale associated with electron cooling. Otherwise, the accelerated proton
configuration may remain self-similar beyond the electron-cooling scale. In this case, the accelerated protons have a
time-dependent energy-cutoff and may (if their spectrum is sufficiently flat) modify the structure of the shock with
time, rendering our assumption of a steady-state inaccurate (self-similarity may still exist). We are not aware of
observational constraints (on the shock thickness, say) that rule out this possibility.

\section{Extensions of the Model}
\label{sec:OtherCases}

In the previous sections we have motivated the self-similarity assumption based on observational evidence that is relevant directly to the downstream of strong, non-magnetized shocks. It is possible, however, that in other related circumstances, in which power-law distributions of energetic particles are interacting with non-magnetized MHD plasmas, similar scalable solutions arise. In \S~\ref{Upstream} we examine the possibility that the hydro-magnetic structure of the upstream is also self-similar. In \S~\ref{Homogenous} we discuss time-dependent, homogenous toy models, in which
energetic particles with a power-law distribution in momentum are added to a plasma.

\subsection{Upstream}
\label{Upstream}

It is expected that the accelerated component precursor generates turbulence in the upstream \citep{Blandford87}. In the
context of GRB's, it was recently claimed that the magnetic field energy density in the upstream is larger than that
typical to the typical ISM by at least 3 orders of magnitude \citep{Zhuo06}, which suggests that high energy particles
generate waves upstream of the shock. If this is indeed the case, it is likely that higher momentum particles are
important at larger distances from the shock, which suggests that the characteristic length scale for variations in the
fields grows with the distance from the shock. Therefore, the EM structure in the upstream may also have a self-similar
character, although we do not have as strong observational evidence for self-similarity in the upstream as we have for
such behavior in the downstream.

Consider the conditions many skin depths upstream of the shock. The equations governing the
accelerated component's distribution function and the EM fields are the same as for the downstream. In particular we expect the relation between the scaling indices to be
\begin{align}\label{ParticleScaling2}
&s_{p1}=s_{B1}+1;~s_{f1}=-2s_{B1}-4,
\end{align}
where the subscript 1 denotes the upstream. As explained in \S~\ref{sec:implications}, assuming that the accelerated
component reaches the shock front, we expect the distribution function to have a power law momentum dependence with a
power law index $l_p=s_f/s_p$. From the same considerations, we must have in the upstream $l_p=s_{f1}/s_{p1}$. We thus
find that the scaling indices of the upstream and downstream are equal. In particular, for the expected value $s_B=0$,
the energy in the magnetic field does not decay as we go farther into the upstream. This result, which cannot be true for
arbitrary distances in the upstream, is related to the diverging energy in accelerated particles. Note also, that the
value of $\bar B$, which is constant according to our analysis in the
upstream and far downstream regions, is expected to change considerably in the shock vicinity (distance of order $l_{sd}$
where the self similarity breaks down), and thus is probably different in the upstream and downstream.

Although the scaling relations are the same for the upstream and the downstream, the accelerated distribution function
can have a qualitatively different spatial dependence.
Note that for sufficiently high momentum the distribution is continuous across the shock.
In particular, for a given momentum, the accelerated particle distribution function in the upstream is expected to decay
exponentially (if the acceleration process is DSA, see e.g. \citealt{Kirk00}), while in the downstream it is expected to
remain approximately constant.

The analysis of the fluid component in the upstream is more complicated, as the magnitude of the (velocity, pressure)
fluctuations and the amount of fluid heating are unknown.

\subsection{Homogenous configurations}
\label{Homogenous}

In this subsection we consider the development of magnetic fields in homogenous configurations as a consequence of the
interaction of the thermal fluid with the accelerated particles (for related simulations see, e.g.,
\citealt{Lucek00,Bell04}). Such configurations are much simpler to simulate and analyze than space-dependent
configurations, and may give insights into dynamically important mechanisms.

If spatial inhomogeneity has an important role in the acceleration mechanism in collisionless shocks (as it does in the
case of Fermi acceleration), it is likely that no acceleration takes place in homogenous configurations. In order to see
the effects of an accelerated component on the dynamics, it would be interesting to perform homogenous simulations with
an accelerated component (with a power-law distribution function) included in the initial conditions.

In the following, we assume a homogenous configuration with a relativistic accelerated component with some anisotropic
distribution function $f_\al(\bp)=g_\al(\mu)p^{l_p},~~p>p_{min}$, included in the initial conditions, where
$\mu=\cos(\bp\cdot\hat z)$ (the $z$ axis is chosen to represent the direction of the shock normal; the cylindrical
symmetry around this axis is present in planar non-magnetized shocks and there is no reason not to include it in
homogenous simulations). We assume that the fluid component may be described by single fluid equations for which there is
no inherent physical length scale. We assume initial neutrality, in the sense that any charge density (electric current)
implied by $f_\al$ is compensated by an opposite charge density (current) carried by the fluid.

Quite generally, a non-isotropic distribution is unstable. The distribution function is expected to evolve from the
initial unstable configuration to a stable (probably isotropic) configuration by the development of instabilities
followed by dissipation. Particles with larger momentum will respond more slowly to the generated EM fields (which are
limited in strength by the initial energy) and on longer length scales (perhaps of the order of their Larmor radius).
Fluctuations of the EM fields on small scales are expected to decay with time \citep{Gruzinov01a} while EM field
generation at later times due to instabilities in larger momentum regimes is expected to occur on longer length scales.
It is thus likely that the magnetic field length scale grows with time. Self similarity is to be expected once the length
scale of the magnetic field is much larger than the Larmor radius of particles having $p_{min}$.

The time development may be
affected by the choice of $g_\al$. In particular, the value of the bulk electric current carried by the initial
population $\propto \int_{-1}^1\mu [g_p(\mu)-g_e(\mu)]$ and bulk velocity $\propto \int_{-1}^1\mu [g_p(\mu)+g_e(\mu)]$
may affect the nature of the instabilities involved. As we show in this subsection, the self similarity assumption (as
long as it is valid) allows determination of some physically interesting quantities without dependence on the precise
instability mechanism.

In this case, we cannot assume a steady-state. However, there is a full (three dimensional) translational symmetry
(homogeneity). We consider averaged quantities at fixed times. In particular, Eq. \eqref{eq:Avaregedf} and Eq.
\eqref{eq:Bcor} are replaced by:
\begin{align}
\ave{f_\al}(\bp,t)\equiv\lim_{r\ra\infty}\inv{\frac43\pi
r^3}\int_{x'<r} d^3\bx' f_\al(\bx+\bx',\bp,t)
\end{align}
and
\begin{align}
&B_{ij}(\Dbx,\De t,t)\equiv\cr &\lim_{r\ra\infty}\inv{\frac43\pi r^3}\int_{x'<r} d^3\bx'
B_i(\bx+\bx',t)B_j(\bx+\Dbx+\bx',t+\De t),\cr
\end{align}
where these functions do not depend on $\bx$ due to the homogeneity ($\bar B$ defined similarly).

We assume that a scaling property, following Eqs. \eqref{eq:slfB}-\eqref{eq:slfCorB}, is obtained at late times, i.e.
\begin{align}
\label{eq:slfBHom} \bar B(L)= \left(\frac{L}{L_0}\right)^{s_B} \bar B(L_0),
\end{align}
\begin{align}
\label{eq:slfFHom} \ave{f_{\al}}(\bp,L)= \left(\frac{L}{L_0}\right)^{s_f}
\ave{f_\al}\left[\frac{\bp}{(L/L_0)^{s_p}},L_0\right],
\end{align}
and
\begin{align}\label{eq:slfCorBHom}
B_{ij}(\Dbx,\Dt,L)=\left(\frac{L}{L_0}\right)^{2s_B} B_{ij}\left[\frac{\Dbx}{L/L_0},\frac{\Dt}{(L/L_0)^{s_t}},L_0\right],
\end{align}
with $L\propto t^{1/s_t}$, $T\propto t$ ($s_t>0$).

An important difference with respect to the shock case arises from the absence of a
 finite bulk velocity,
and therefore the assumption that time scales as distance no longer applies in general. We study cases in which the
magnetic fields are stronger than electric fields and thus restrict $s_t\geq1$. A value of $s_t>1$ is consistent with
Vlasov's equation, Eq.~\eqref{eq:VlasovRel}, if we neglect the time derivative term and the electric field term, which is
justified for $s_t>1$.

The distribution function of the accelerated component is by definition finite at $t\ra0$ (also at $L\ra0$). In analogy
to Eq. \eqref{eq:highp}, we therefore have $\ave{f_{\al}}(\bp,t)\propto p^{s_f/s_p}$ for $p\ra\infty$. On the other hand,
$\ave{f_{\al}}(\bp,t=0)\propto p^{l_p}$, which implies that $s_f/s_p=l_p$ (this also shows that a finite initial
distribution must be a power-law in order to achieve self-similarity). From Vlasov's equation (and the expression for the
electric current) we find, as for the shock case, that
\begin{align}
&s_p=s_B+1, \quad s_f=-2s_B-4.
\end{align}
Here it is useful to solve for the scaling indices in terms of the initial power-law index $l_p$,
\begin{align}\label{ParticleScalingH}
s_B=-\frac{l_p+4}{l_p+2},~s_p=-\frac{2}{l_p+2},~s_f=-\frac{2l_p}{l_p+2}.
\end{align}

Assuming that the fluid can be considered as infinitely conducting (see \S~\ref{sec:InfiniteCon}), we have $u\sim L/T$ so
$s_u=1-s_t$. Assuming further that the magnetic force is not negligible in the fluid momentum equation
\eqref{eq:FluidMomentumEquation} we find $s_u=s_B$ (this is reasonable as the energy source is the energetic component,
and energy is transferred to the fluid through the magnetic fields). Together with Eq. \eqref{ParticleScalingH} we have
\begin{align}\label{TimeScalingH}
s_t=1-s_B=\frac{2(l_p+3)}{l_p+2}.
\end{align}
In particular this implies
$\bar B\propto L^{s_B}\propto t^{s_B/s_t}=t^{-(l_p+4)/2(l_p+3)}$.

\section{Some general implications}
\label{sec:general}

If self-similarity holds, it has important implications for any model of particle acceleration and/or field generation.
In \S~\ref{Medvedev} we show that a previously suggested model, that describes the flow in terms of merging current
filaments \citep{Medvedev05} in a self similar manner, must be generalized in order to be a self-consistent model. We
show that after the appropriate generalization the model follows the scaling relations derived in \S~\ref{Homogenous},
and that the generalization significantly modifies the model's predictions. In \S~\ref{ApplicationDSA} we discuss the
relevance of our analysis to DSA.

\subsection{Current merger model}
\label{Medvedev}

\citet{Medvedev05} suggest a model of coalescing electric current filaments in a quasi two-dimensional configuration for
describing the magnetic field evolution in the downstream of collisionless shocks. In this model there is no distinction
between accelerated and thermal particles, so it does not hold information regarding the energy dependence of the
particle distribution function. As the model is essentially homogenous and self-similar, it is interesting to compare it
to the scalings derived in \S~\ref{Homogenous}.

This model assumes that electromagnetic instabilities in the shock lead to the formation of current filaments, which may
be approximated as infinite in length and having equal electric currents oriented parallel to the flow, half of the
currents positive (oriented along the flow) and half negative. The filaments are assumed to be distributed randomly in
space. This simple model assumes that current filament coalescence evolves in discrete steps. In each step the filaments
carrying similarly oriented currents, are divided into neighboring pairs, which attract each other and merge. Two
filaments with diameter $D$, inertial mass per unit length $\mu$, and current $I$, unite to form a new filament of
diameter $D'=\sqrt{2}D$ (conserving area), mass per unit length $\mu'=2\mu$, and current $I'=2I$. In each step, the
typical distance $d$ between neighbor filaments with similarly oriented currents, grows by a factor of $\sqrt{2}$,
$d'=\sqrt{2}d$, as a consequence of the reduced filament number density. It is assumed that $d\simeq 2D$ which implies
that the filaments roughly fill the space with their area. Initially, all the filaments are identical (except for the
sign of the current), and are hence identical at any given time. The coalescence time, here denoted by $\tau_{coa}$, is
estimated by analyzing two isolated, parallel, infinite current filaments attracting each other.

The fact that $D$ and $d$ change in each merger by the same factor, implies that the configuration following each merger
is identical to that before the merger, with re-scaled parameters. In particular, the filament coalescence time,
$\tau_{coa}\propto\mu^{1/2}cD/I$, and the maximal velocity during coalescence, $\trm{v}\propto I/(c\mu^{1/2})$, scale as
$\tau_{coa}'=\tau_{coa}$ and $\trm{v}'=\sqrt{2}\trm{v}$. The fact that the filament velocity grows with time, led the
authors to the conclusion, that velocities of the order of the speed of light will be reached. Once the filaments move
with $\trm{v}\sim c$, the coalescence temporal behavior changes considerably.

This suggested merger model is problematic, as it implies that the magnetic field grows by a factor of $\sqrt{2}$ in each
step. For example, the magnetic field produced by a single filament at its edge,
$B=4I/cD$, scales as $B'=\sqrt{2}B$. More generally, the coalescence suggested is self-similar with
$\textbf{j}'(\textbf{x})=\textbf{j}(\textbf{x}/\sqrt{2}),$ implying
$\textbf{B}'(\textbf{x})=\sqrt{2}\textbf{B}(\textbf{x}/\sqrt{2})$. To see this, note that the current density carried by
each filament, $j\sim I/D^2$, does not change in the coalescence, while the filament diameter and distance between
filaments grow by $\sqrt{2}$. The fact that the magnetic field grows by a constant factor in each step, implies that
after a few steps the magnetic energy density grows beyond equipartition, which is impossible. This can be corrected by
changing appropriately the coalescence conditions.

Since the physical process of the merger is complicated, there is no simple way to determine the post-merger current
directly. In general, one should therefore set the current of the merged filament to $(\sqrt{2})^\zeta I$ where $\zeta$
is a parameter of the model (the mass per unit length is set to $2\mu$ since mass is conserved). In this case, the
coalescence is self-similar with $\textbf{j}'(\textbf{x})=(\sqrt{2})^{\zeta-2}\textbf{j}(\textbf{x}/\sqrt{2}),$ implying
that $\textbf{B}'(\textbf{x})=(\sqrt{2})^{\zeta-1}\textbf{B}(\textbf{x}/\sqrt{2})$ and the magnetic field amplitude does
not grow for $\zeta\leq1$ .
The scaling of the filament coalescence time and maximal velocity changes to $\tau_{coa}'=(\sqrt{2})^{2-\zeta}\tau_{coa}$
and $\trm{v}'=(\sqrt{2})^{\zeta-1}\trm{v}$.

The (necessary) generalization of the current merger model
strongly influences the conclusions that can be drawn based on
this model. The growth in length scale is only determined up to a
free parameter $\zeta$. In order for the magnetic field not to
diverge with time, we must have $\zeta\leq1$  (if the magnetic
field amplitude is postulated to be constant in time, $\zeta=1$).
This implies that the velocities do not grow (remain constant for
$\zeta=1$) and therefore do not reach the speed of light.

The (corrected) scalings are compatible with the scalings for the fluid derived in \S~\ref{Homogenous}, with
$s_B=\zeta-1, s_t=2-\zeta, s_u=\zeta-1$. Note, that the self-similar analysis allows us to reach most of the physically
interesting conclusions without making oversimplifying and model-specific assumptions, such as the calculation of the
length of the time step using two isolated filaments in the current merger model.

It has been claimed \citep{Milosavljevic05a} that current filaments are unstable to kink-like modes and that the two
dimensional configuration is therefore disrupted.
We note that even if the instability is efficient, this does not rule out the model.
Since the instabilities are derived within the framework of MHD, for which the above scalings hold, the perturbation's
growth rate scales as all other time scales, i.e. $\tau_{ins}'=(\sqrt{2})^{2-\zeta}\tau_{ins}$ where $\tau_{ins}$ is the
inverse growth rate. If initially $\tau_{ins}>\tau_{coa}$ (which is not ruled out), the filaments merge before they are
destroyed by the instabilities, and this holds, i.e. $\tau_{ins}>\tau_{coa}$, for all subsequent mergers.

\subsection{Diffusive shock acceleration}
\label{ApplicationDSA}

Diffusive shock acceleration (DSA) is the mechanism believed to be responsible for the production
of non-thermal populations of charged particles in collisionless shocks. In this first-order Fermi acceleration process,
particles gain energy by repeatedly bouncing between the converging upstream and downstream flows. In the lack of a
self-consistent theory for the interaction between electromagnetic waves and accelerated particles in collisionless
shocks, most progress was made under the "test-particle" approximation. In this approach, the effects of the waves are
modelled by some particle-scattering Ansatz, while the influence of the particles on the waves and on the shock structure
are neglected (for reviews, see \citealt{Drury83, Blandford87, Malkov01}).

In the case of non-relativistic shocks, DSA has been successful in reproducing the power-law spectra observed in strong
shocks, under very general assumptions regarding the scattering mechanism \citep{Krymskii77, Axford77, Bell78,
Blandford78}. The analysis of relativistic shocks is more complicated, mainly because the non-thermal particle
distribution is not isotropic. The particle spectrum was calculated in relativistic shocks under various assumptions
regarding the scattering mechanism, using monte-carlo simulations [e.g. \citet{Ellison90, Ostrowski91}], and by numerical
\citep{Kirk87, Gallant99, Vietri03, Blasi05} or analytical \citep{Keshet05} study of the transport equations. In general,
a power-law distribution function of accelerated particles is found, with indices that depend on details of the model and
usually satisfy $l_p<-4$. In the case of ultra-relativistic shocks with isotropic, small-angle scattering, numerical
studies have converged on a spectral index $l_p\simeq-4.22$ \citep{Bednarz98,Kirk00, Achterberg01}, in agreement with GRB
afterglow observations and with the analytic result $l_p=38/9$ \citep{Keshet05}. However, this value was demonstrated
numerically \citep{Ballard92, Ostrowski93, Ellison02, Ellison04, Meli03a, Meli03b, Bednarz04, Niemiec04,
Lemoine03,Lemoine06} and analytically \citep{Keshet05} to be sensitive to the scattering mechanism, which is poorly
constrained.

In our analysis, the fluctuations in the fluid provide a natural scattering mechanism for the accelerated particles.
Since the analysis reflects near equipartition between fluid fluctuations, accelerated particles and magnetic fields, it
does not naturally evoke (but does not rule out) a test-particle approach. A spectral index $l_p\neq4$ can be reconciled
with our analysis only if we relax some of our assumptions. For example, if we assume that the magnetic field energy is
constant with distance from the shock, $l_p<-4$ would imply that the currents carried by the accelerated component
$\bj_h$ decrease faster than the total current $\bj\sim \nabla\X\bB$, and thus the effect of the accelerated particles on
the magnetic field is negligible at large distances (in this case, the test particle assumption is self-consistent). If,
on the other hand, we assume that the magnetic field decreases with distance as $s_B=-(l_p+4)/(l_p+2)$ (see
\S~\ref{sec:implications}), a value $l_p<-4$ would contradict our assumptions regarding the self similarity of the fluid
component (for example, the magnetic force would become negligible in the momentum conservation equation, or the fluid
would no longer be highly conductive and its currents would become negligible).

The assumption of self-similarity places constraints on some properties of DSA. As an illustration, consider the case of
small-angle scattering, parameterized by a propagation-angle diffusion coefficient $D_{\mu\mu}$. The stationary transport
equation can be written as \citep{Kirk87}:
\begin{align} \label{eq:DSATransport}
\gamma_i(\trm{v}_i/c+\mu_i)\pr_{z_s}\ave{f_i}(p_i,\mu_i,z) =
\pr_{\mu_i}\left[D_{\mu\mu}(\mu_i,p_i,z)\pr_{\mu_i}\ave{f_i}\right]
\end{align}
where subscript $i=1,2$ denotes upstream or downstream parameters, respectively, $\gamma_i=(1-\trm{v}_i^2/c^2)^{-1/2}$,
and $\mu=\cos(\bp_i\cdot\hbz)$. The momentum $\bp_i$ (and therefor also $\mu$) is measured in the fluid rest frame,
whereas $z$ is measured in the shock frame. With boundary conditions $\ave{f_i}(p_i,\mu_i,-\infty)=0$,
$\ave{f_i}(p_i,\mu_i,\infty)=\ave{f_i}_{\infty}(p_i)$ and $\ave{f_1}(p_1,\mu_1,0)=\ave{f_2}(p_2,\mu_2,0)$,
 where
$(p_1,\mu_1)$ and $(p_2,\mu_2)$ are related through an appropriate Lorenz boost. The solution is unique up to the
normalization of $\ave{f_i}$ \citep{Kirk00}.

The small-angle scattering assumption requires that the Larmor radius of the particles $R_L$ is much larger than the
magnetic field length-scale $L$. In our framework, for any given momentum $p$ this is true up to a limited distance from
the shock (where $L<pc/e\bar B$). Consistency of DSA theory thus requires that $\ave{f_i}(p_i,\mu_i,z)$ converge to its
limiting value $\ave{f_i}_{\infty}(p_i)$ within this range.

In order to comply with the self similar scaling of Eq.~(\ref{eq:slfF}), we must require that
\begin{align}\label{scalingDSA}
\ave{f_i}(p_i,\mu_i,z)=\left(\frac{z}{z_0}\right)^{s_f}
\ave{f_i}\left[\frac{p_i}{(z/z_0)^{s_p}},\mu_i,z_0\right].
\end{align}
This scaling can be reconciled with the transport equation [Eq.~(\ref{eq:DSATransport})] if and only if the diffusion
function scales as
\begin{align}\label{ConditionDSA}
D_{\mu\mu}(p_i,\mu_i,z)=\frac{z_0}{z}D_{\mu\mu}\left[\frac{p_i}{(z/z_0)^{s_p}},\mu_i,z_0\right].
\end{align}
The boundary conditions are invariant to the above scaling (at the shock front $p_1\propto p_2$). Under this scaling,
$\ave{f_i}(p_i,\mu,z)\propto p^{l_p}$ at $z=\pm \infty$, where $l_p=s_f/s_p$. The values of $s_f$ and $l_p$ are
determined in a non-trivial way by the function $D_{\mu\mu}(p_i,\mu_i,z)$.
If $s_p=1$, as is expected when the plasma is highly conductive, we may write Eq.~(\ref{ConditionDSA}) as
\begin{align}\label{ConditionDSA2}
D_{\mu\mu}(p_i,\mu_i,z)= z^{-1} \,\tilde{D}_{\mu\mu}(\mu_i,p_i/z).
\end{align}
As an example of the above scaling, consider a highly relativistic particle scattered by weak magnetic fluctuations of
correlation length $L\ll R_L$. The particle trajectory may then be described as a random walk in $\mu$, and the resulting
diffusion function is $D_{\mu\mu}(p,\mu,z)=D(\mu)L R_L^{-2}$. For $L\propto z$ (and $s_B=0$), we obtain
$D_{\mu\mu}(p,\mu,z) \propto z^{-1} (p/z)^{-2}$.

\section{Discussion}
\label{sec:Discussion}

We have studied the consequences of the assumption that the downstream flow of non-magnetized, collisionless shocks is
self-similar. This assumption is motivated by the existence of a strong magnetic field many skin depths downstream of the
shock front, as inferred from observations of GRB afterglows and young SNR's. This suggests that the correlation length
of magnetic field fluctuations, $L$, diverges with the distance $D$ from the shock front. Although our analysis was
motivated by evidence for self-similarity in GRB afterglows, and possibly also in SNRs, it may apply to other systems
with similar characteristics, such as the shocks involved in the large scale structure of the Universe.

As the electromagnetic fields and particle distributions at any given point
fluctuate with time, a stationary shock structure
may be described
by the averages (over planes parallel to the shock front) and correlation functions of the fluctuating quantities, which
depend only on the distance from the shock (see \S~\ref{sec:stationary}). The self-similarity assumption implies that
$L\propto D$ and that the averages and correlation functions at different distances $D$ from the shock, corresponding to
different values of $L$, are related to each other by simple scaling transformations (see \S~\ref{sec:DownStream}), e.g.
\begin{align}
\nonumber \bar B(L)= \left(\frac{L}{L_0}\right)^{s_B} \bar
B(L_0),
\end{align}
\begin{align}
\nonumber \ave{f_{\al}}(\bp,L)=
\left(\frac{L}{L_0}\right)^{s_f}
\ave{f_\al}\left[\frac{\bp}{(L/L_0)^{s_p}},L_0\right],
\end{align}
\begin{align}
\nonumber B_{ij}(\Dbx,\Dt,L)=\left(\frac{L}{L_0}\right)^{2s_B}
B_{ij}\left[\frac{\Dbx}{L/L_0},\frac{\Dt}{(L/L_0)^{s_t}},L_0\right],
\end{align}
where the scaling exponent of the magnetic field must satisfy
\begin{equation}
\nonumber    -1<s_B\le0.
\end{equation}
A schematic illustration of a self similar downstream configuration is presented in figure \ref{fig:Illustration}.

We have argued (see \S~\ref{sec:2comp}) that the similarity assumption suggests that the plasma may be approximately
described as a combination of two self-similar components: a kinetic component of energetic particles, and an MHD-like
component representing "thermal" particles. We have argued that the energetic particles are likely to carry a significant
fraction of the current, and derived (using the Maxwell-Vlasov equations) the scaling of the characteristic time for
variations in the physical quantities, the scaling of the characteristic particle momentum and the scaling of the
particle distribution function normalization:
\begin{equation}
\nonumber s_t=1,\quad s_p=s_B+1,\quad s_f=-4-2s_B.
\end{equation}
These relations imply that the characteristic Larmor radius of energetic particles scales as $L$, and that the energy
density of energetic particles in any momentum interval, with the interval scaling as $L^{s_p}$, scales as the magnetic
field energy density $\propto L^{2s_B}$. We have then shown (see \S~\ref{sec:implications}) that (under the assumption
that accelerated particles reach the shock front and/or are advected to the downstream) the spectrum of accelerated
particles is
\begin{equation}
\nonumber    dn/dE\propto E^{-2/(s_B+1)},
\end{equation}
and that the scaling of the magnetic correlation function (for $\Dx\ra\infty$) is
\begin{equation}
\nonumber   \ave{B_i(\bx)B_j(\bx+\Dbx)}\propto\Dbx^{2s_B}.
\end{equation}
Similar conclusions can be drawn regarding various other correlation functions.

The "thermal" particles were discussed in \S~\ref{sec:fluid}. In \S~\ref{sec:InfiniteCon} we have argued that the thermal
component may be considered as an infinitely conducting fluid. We have shown that in this case $s_B=0$ and the scalings
are completely determined, e.g. $dn/dE\propto E^{-2}$ and $B\propto D^0$, with possible logarithmic corrections. We have
derived in \S~\ref{sec:fluidEqs} a closed set of equations for the fluid component, under the simplifying assumptions of
an ideal fluid and neglecting corrections of order $\varepsilon_B$.

The self similarity assumption and its implications do not hold for arbitrarily large distances and high particle momenta
where new physical processes that were not taken into account in our analysis come in to play. For example, our
assumption of an infinite planar shock is invalid at distance scales of the order of the blast wave radius, and the use
of the Vlasov equation is invalid for high momenta where radiative effects cannot be ignored. Upper cutoffs to the
distance scale and the momentum range described by the self-similar solution, or logarithmic corrections to this
solution, are possible remedies of the energy divergence when $s_B=0$, as discussed in \S~\ref{sec:divEng}.

If self-similarity holds, it has important implications for any model of particle acceleration and/or field generation.
In \S~\ref{ApplicationDSA} we have shown that the velocity-angle diffusion coefficient for small-angle scattering in
diffusive shock acceleration models must satisfy $D_{\mu\mu}(\bp,D)=D^{-1}\tilde{D}_{\mu\mu}(\bp/D)$ (where $\bp$ is the
particle momentum).

In \S~\ref{Medvedev} we have discussed the model suggested by \citet{Medvedev05} for the generation of a large scale
magnetic field through hierarchical merger of current-filaments. We have shown that in order to avoid a diverging
magnetic field, the model must be generalized by allowing a more general scaling of the electric current of merged
filaments. The generalized model follows the scaling laws we have derived in \S~\ref{Homogenous}. This implies that the
self-similar analysis allows us to reach most of the physically interesting conclusions without making oversimplifying
and model-specific assumptions (such as those related to the calculation of the merger time). The predictions of the
generalized model differ substantially from those of the original model. It predicts, e.g., a scale independent merging
velocity, rather than an increasing velocity approaching the speed of light (this is valid for a non-decaying magnetic
field; for a decaying field, the velocity decreases with scale). Finally, we have shown that the instability of current
filaments \citep[which was pointed out by][]{Milosavljevic05a} does not necessarily imply that the current merger model
is not viable.

In \S~\ref{Homogenous} we have pointed out that the
self-similarity assumptions may be tested through their
predictions for the evolution of homogenous (time-dependent)
plasmas, which may be accessible to direct numerical simulations.
An inclusion (at the initial conditions) of an anisotropic,
power-law spectrum of high energy particles, $dn/dE\propto
E^{2+l_p}$, in homogenous simulations may lead to a self-similar
evolution in time, described by the scaling exponents given in
Eq.~(\ref{ParticleScalingH}). Assuming that the fluid can be
considered as infinitely conductive, we predict that the magnetic
field evolution follows $B\propto t^{-(l_p+4)/2(l_p+3)}$.

Our self-similar model predicts the scaling of all physical quantities related to the accelerated  particles,
electromagnetic fields and fluid fluctuations. In particular, the particle spectrum is related to the magnetic field
scaling: An accelerated particle distribution with a spectral index $l_p<-4$ indicates a decay of the magnetic field
amplitude with distance from the shock according to $D^{-(l_p+4)/(l_p+2)}$. Such a decay of the magnetic field may be
detectable, for example through the spatial dependence of synchrotron emission from non-thermal electrons gyrating in the
downstream magnetic fields.

Finally, we should emphasize two major open questions related to our analysis, which need to be resolved. First, the
validity of the description of the "thermal" component in terms of single-fluid MHD equations requires verification. Second, the diverging energy in accelerated particles that results from the
distribution $n(E)\propto E^{-2}$, indicates that our assumptions cannot be valid for arbitrarily high momentum (see
discussion and possible remedies at the end of \S~\ref{sec:implications}). A similar divergence is identified in the energy of magnetic fields and fluid fluctuations.

\acknowledgements This work was supported in part by ISF and AEC
grants. We thank A. Gruzinov, A. Loeb \& A. Sagiv for helpful
discussions.

\appendix

\section{Collisionless Shocks in 3D Particle-in-cell simulations}\label{appendix:PICs}

In recent years, particle in cell (PIC) simulations have provided important preliminary clues to the nature of
collisionless shocks. Fully three-dimensional PIC simulations have began to explore aspects of such shocks in
non-homogeneous \citep{Nishikawa03,Frederiksen04,Spitkovsky05} and homogeneous \citep{Silva03,Jaroschek04,Romanov04}
flows. The main conclusions relevant for collisionless shocks are:

\begin{itemize}

\item Confirmation that shocks in pair-plasma (with no highly
accelerated particles) are mediated by electromagnetic, two-stream (Weibel-like) instabilities if the upstream is
sufficiently non-magnetized (magnetic energy less than $\sim1\%$ of the kinetic energy; \citealt{Spitkovsky05}) and cold.
In interpenetrating shells, this instability produces filaments in density (if $\Gamma\gtrsim5$) and in electric current
\citep{Jaroschek05}, initially with $\lesssim$skin depth thickness, and transverse (perpendicular to the flow) magnetic
fields. Electrons in ion-electron plasma behave similarly \citep{Frederiksen04}.

\item Preliminary evidence for the existence of relativistic,
collisionless shocks in a non-magnetized pair-plasma. \citet{Spitkovsky05} resolves a shock-like behavior in a 3D
simulation of interpenetrating pair-plasma shells, each with Lorentz factor $\Gamma=15$. A density jump in approximate
agreement with the Rankine-Hugoniot adiabat extends over $\sim70l_{sd}$, and coincides with a region of
near-equipartition magnetic fields. Here, $l_{sd}=c/\omega_{p}$ and $\omega_{p}=(4\pi ne^{2}/\Gamma m)^{1/2}$ are the
skin-depth and plasma frequency, respectively, of simulated particles of mass $m$, number density $n$ and Lorentz factor
$\Gamma$.

\item Possible indications for self-similarity. In
interpenetrating plasma shells, the filaments merge after saturation and grow in a hierarchical process, maintaining a
configuration of similar features on gradually larger scales, with magnetic fields decaying slowly and possibly
saturating at $\epsilon_{B}\lesssim1\%$ [in pair plasmas; \citet{Silva03,Medvedev05}]. Frederiksen et al. (2004; Figure
3) provide evidence that the power-spectrum of transverse magnetic fields is roughly a power-law, $P_{k}\propto
k^{-\alpha}$ with $\alpha\simeq2.3-3.0$, extending to gradually larger scales.
\end{itemize}

Present 3D PIC simulations are limited to small simulation-box volumes, $V<10^{6}l_{sd}^{3}$, and to short simulation
durations, $T<10^{3}\omega_{p}^{-1}$. Here, $\omega_{p}$ and $l_{sd}$ corresponds to the species with fastest-response;
simulating an ion-electron plasma implies much smaller volumes and durations when written in terms of the ion parameters.
In order to resolve any effects associated with the ions, an effective, small proton to electron mass ratio,
$\tilde{m}_p/m_e\lesssim 20$ (with present computational resources) must be used, and the preliminary results thus
obtained are not easily extrapolated to more realistic mass ratios. These constraints limit the relevance of any
conclusion drawn from such simulations, and may even undermine their reliability:

\begin{itemize}

\item The length of the simulation box parallel to the flow is too
small in most cases to resolve a shock , even in pair-plasma [with the possible exception of \citealt{Spitkovsky05}]. The
configuration found in most simulations thus represents merely a transient stage in the formation of a shock, which may
bare little relevance to its steady-state. Moreover, the short length of the box suppresses longitudinal modes of long
wavelengths, thus distorting the plasma evolution by effectively reducing it to 2D. For example, an exponential decay of
the magnetic energy (after growth saturation) abruptly stops in simulations \citep{Silva03, Jaroschek05}, possibly due to
this effect.

\item The small size of the simulation box ($\lesssim$ tens of
skin-depths) perpendicular to the flow places an artificial cutoff on transverse modes with long wavelengths. Simulations
indicate that the transverse scale of the most energetic modes grows rapidly, reaching the box size cutoff at early
stages of the simulation (\citealt{Silva03}, Figure 3; \citealt{Frederiksen04}, Figures 3-4). After this occurs, the
simulations are strongly affected by the boundary conditions, and the simulated evolution is probably highly distorted.

\item The Larmor radius of particles of Lorentz factor $\gamma$ is
$R_{L} = l_{sd} [\epsilon_{B}(\gamma^2-1)/2\Gamma(\Gamma-1)]^{-1/2}$, where $\Gamma$ is the average Lorentz factor. A
simulation box tens of skin-depths long is therefor not sufficiently large to simulate particle magnetization when the
ratio between magnetic and thermal energy densities is $\epsilon_{B}\lesssim1\%$. Moreover, the box is much too small to
resolve acceleration of particles to high energies, so any Fermi-like acceleration process is suppressed. Electrons are
observed to be accelerated in ion-electron simulations \citep{Nishikawa03,Hededal04}, but to energies smaller than
$\Gamma m_p c^2$.

\end{itemize}

\bibliographystyle{apj}

\begin{thebibliography}{21}
\expandafter\ifx\csname
natexlab\endcsname\relax\def\natexlab#1{#1}\fi
\bibitem[{{Achterberg} {et~al.}(2001){Achterberg}, {Gallant}, {Kirk}, \&
  {Guthmann}}]{Achterberg01}
{Achterberg}, A., {Gallant}, Y.~A., {Kirk}, J.~G., \& {Guthmann},
A.~W. 2001,
  \mnras, 328, 393

\bibitem[{{Axford} {et~al.}(1977)}]{Axford77}
{Axford}, W. I. , {Leer}, E. \& {Skadron}, G. 1977, Proc. 15th Int. Cosmic Ray Conf., Plovdiv (Budapest: Central Research
Institute for Physics), 11, 132

\bibitem[{{Ballard} \& {Heavens}(1992)}]{Ballard92}
{Ballard}, K.~R. \& {Heavens}, A.~F. 1992, \mnras, 259, 89

\bibitem[{{Bamba} {et~al.}(2003)}]{Bamba03}
{Bamba}, A., {Yamazaki}, R., {Ueno}, M., \& {Koyama}, K. 2003,
\apj, 589, 827

\bibitem[{{Bednarz}(2004)}]{Bednarz04}
{Bednarz}, J. 2004, \pasj, 56, 923

\bibitem[{{Bednarz} \& {Ostrowski}(1998)}]{Bednarz98}
{Bednarz}, J. \& {Ostrowski}, M. 1998, Physical Review Letters, 80,
3911

\bibitem[{{Begelman} {et~al.}(1994)}]{Begelman94}
{Begelman}, M.~C., {Rees}, M.~J.,  \& {Sikora}, M. 1994, \apj,
429, L57 (1994);

\bibitem[{{Bell}(1978)}]{Bell78}
{Bell}, A. R. 1978, \mnras, 182, 147

\bibitem[{{Bell}(2004)}]{Bell04}
{Bell}, A.~R. 2004, \mnras, 353, 550

\bibitem[{{Bell}(2005)}]{Bell05}
{Bell}, A.~R. 2005, \mnras, 358, 181

\bibitem[Berger et al.(2003)]{Berger03} Berger, E., Kulkarni,
S.~R., \& Frail, D.~A.\ 2003, \apj, 590, 379

\bibitem[{{Blandford} \& {Ostriker}(1978)}]{Blandford78}
{Blandford}, R. D. \& {Ostriker} J. 1978, \apj, 221, L29

\bibitem[{{Blandford} \& {Eichler}(1987)}]{Blandford87}
{Blandford}, R. \& {Eichler}, D. 1987, \physrep, 154, 1

\bibitem[{{Blandford} \& {McKee}(1976)}]{Blandford76}
{Blandford}, R. D. \& McKee, C. F. 1976, Phys. Fluids, 19, 1130

\bibitem[{{Blasi} \& {Vietri}(2005)}]{Blasi05}
{Blasi}, P. \& {Vietri}, M. 2005, \apj, 626, 877

\bibitem[{{Dieckmann} {et~al}(2006)}]{Dieckmann06}
{Dieckmann}, M. E., {Shukla}, P. K. \& {Drury}, L. O. C. 2006,
\mnras, 367,1072

\bibitem[{{Drury}(1983)}]{Drury83}
{Drury}, L.~O. 1983, Reports of Progress in Physics, 46, 973

\bibitem[{{Drury} {et~al.}(1989)}]{Drury89}
{Drury}, L. O'C., {Markiewicz}, W. J., \& {V{\"o}lk}, H.~J. 1989,
\aap, 225, 179

\bibitem[{{Eichler} \& {Waxman}(2005)}]{Eichler05}
{Eichler}, D. \& {Waxman}, E. 2005, \apj, 627, 861

\bibitem[{{Ellison} \& {Double}(2002)}]{Ellison02}
{Ellison}, D.~C. \& {Double}, G.~P. 2002, Astroparticle Physics, 18,
213

\bibitem[{{Ellison} \& {Double}(2004)}]{Ellison04}
---. 2004, Astroparticle Physics, 22, 323

\bibitem[{{Ellison} {et~al.}(1990){Ellison}, {Reynolds}, \&
  {Jones}}]{Ellison90}
{Ellison}, D.~C., {Reynolds}, S.~P., \& {Jones}, F.~C. 1990, \apj,
360, 702

\bibitem[Frail,~Waxman~\&~Kulkarni (2000)]{FWK00}
  Frail, D. A., Waxman, E., \& Kulkarni, S. R. 2000, \apj, 537, 191

\bibitem[Frail et al. (2001)]{Frail01}
  D. A. Frail et al. 2001, \apj, 562, L55

\bibitem[{{Frederiksen} {et~al.}(2004){Frederiksen}, {Hededal}, {Haugb{\o}lle},
  \& {Nordlund}}]{Frederiksen04}
{Frederiksen}, J.~T., {Hededal}, C.~B., {Haugb{\o}lle}, T., \&
{Nordlund},
  {\AA}. 2004, \apjl, 608, L13

\bibitem[Freedman \& Waxman(2001)]{Freedman01} Freedman, D.~L., \&
Waxman, E.\ 2001, \apj, 547, 922

\bibitem[{{Gallant} \& {Achterberg}(1999)}]{Gallant99}
{Gallant}, Y.~A. \& {Achterberg}, A. 1999, \mnras, 305, L6

\bibitem[{{Gruzinov}(2000)}]{Gruzinov00}
Gruzinov, A. 2000,
arXiv:astro-ph/0012364

\bibitem[{{Gruzinov}(2001a)}]{Gruzinov01a}
{Gruzinov}, A. 2001a, \apjl, 563, L15

\bibitem[{{Gruzinov}(2001b)}]{Gruzinov01b}
  ---. 2001b,
  arXiv:astro-ph/0111321

\bibitem[Gruzinov \& Waxman (1999)]{Gruzinov99}
 Gruzinov, A. \& Waxman, E.1999, \apj, 511, 852

\bibitem[{{Hededal} {et~al.}(2004)}]{Hededal04}
{Hededal}, C. B., {Haugb{\o}lle}, T., {Frederiksen}, J. T. \&
{Nordlund}, {\AA} 2004, \apj, 617, L107

\bibitem[{{Jaroschek} {et~al.}(2004){Jaroschek}, {Lesch}, \&
  {Treumann}}]{Jaroschek04}
{Jaroschek}, C.~H., {Lesch}, H., \& {Treumann}, R.~A. 2004, \apj,
616, 1065

\bibitem[{{Jaroschek} {et~al.}(2005){Jaroschek}, {Lesch}, \&
  {Treumann}}]{Jaroschek05}
---. 2005, \apj, 618, 822

\bibitem[Kadanoff et al.(1967)]{Kadanoff67} Kadanoff, L.~P., et
al.\ 1967, Reviews of Modern Physics, 39, 395

\bibitem[{{Kato}(2005)}]{Kato05}
Kato, T. N. 2005, Phys. Plasmas 12, 080705

\bibitem[{{Keshet} {et~al.}(2004){Keshet}, {Waxman}, \& {Loeb}}]{Keshet04}
{Keshet}, U., {Waxman}, E., \& {Loeb}, A. 2004, \apj, 617, 281

\bibitem[{{Keshet} \& {Waxman}(2005)}]{Keshet05}
{Keshet}, U. \& {Waxman}, E. 2005, Physical Review Letters, 94,
111102

\bibitem[{{Kirk} {et~al.}(2000){Kirk}, {Guthmann}, {Gallant}, \&
  {Achterberg}}]{Kirk00}
{Kirk}, J.~G., {Guthmann}, A.~W., {Gallant}, Y.~A., \& {Achterberg},
A. 2000,
  \apj, 542, 235

\bibitem[{{Kirk} \& {Schneider}(1987)}]{Kirk87}
{Kirk}, J.~G. \& {Schneider}, P. 1987, \apj, 315, 425

\bibitem[{{Krall}(1997)}]{Krall97}
{Krall}, N.~A. 1997, Advances in Space Research, 20, 715

\bibitem[{{Krymskii}(1977)}]{Krymskii77}
{Krymskii}, G. F. 1977, Dokl. Akad. Nauk SSSR, 234, 1306

\bibitem[{{Lemoine} \& {Pelletier}(2003)}]{Lemoine03}
{Lemoine}, M. \& {Pelletier}, G. 2003, \apjl, 589, L73

\bibitem[{{Lemoine} \& {Revenu}(2006)}]{Lemoine06}
{Lemoine}, M. \& {Revenu}, B. 2006, \mnras, 366, 635

\bibitem[Lucek \& Bell(2000)]{Lucek00} Lucek, S.~G., \& Bell,
A.~R.\ 2000, \mnras, 314, 65

\bibitem[{Li \& Waxman(2006)}]{Zhuo06}
Li, Z. \& Waxman, E. 2006, arXiv:astro-ph/0603427

\bibitem[{{Loeb} \& {Waxman}(2000)}]{Loeb00}
{Loeb}, A. \& {Waxman}, E. 2000, Nature, 405, 156

\bibitem[{Lyubarsky \& Eichler(2005)}]{Lyubarsky05}
Lyubarsky, Y. \& Eichler, D. 2005, arXiv:astro-ph/0512579

\bibitem[{{Malkov} \& {Drury}(2001)}]{Malkov01}
{Malkov}, M. A. \& {Drury}, L. O'C. 2001, Rep. Prog. Phys. 64 429

\bibitem[{{Maraschi}(2003)}]{Maraschi03}
{Maraschi}, L. 2003, in AGNs: from Central Engine to Host Galaxy,
Eds. S. Collin, F. Combes \& I. Shlosman. ASP, Conference Series,
{\bf 290} 275 (2003)

\bibitem[{{Medvedev} {et~al.}(2005){Medvedev}, {Fiore}, {Fonseca}, {Silva}, \&
  {Mori}}]{Medvedev05}
{Medvedev}, M.~V., {Fiore}, M., {Fonseca}, R.~A., {Silva}, L.~O., \&
{Mori},
  W.~B. 2005, \apjl, 618, L75

\bibitem[{{Medvedev} \& {Loeb}(1999)}]{Medvedev99}
{Medvedev}, M.~V. \& {Loeb}, A. 1999, \apj, 526, 697

\bibitem[{{Meli} \& {Quenby}(2003{\natexlab{a}})}]{Meli03b}
{Meli}, A. \& {Quenby}, J.~J. 2003{\natexlab{a}}, Astroparticle
Physics

\bibitem[{{Meli} \& {Quenby}(2003{\natexlab{b}})}]{Meli03a}
---. 2003{\natexlab{b}}, Astroparticle Physics

\bibitem[{Milosavljevic \& Nakar(2005a)}]{Milosavljevic05a}
Milosavljevic, M. \& Nakar, E. 2005a, arXiv:astro-ph/0508464

\bibitem[{Milosavljevic \& Nakar(2005b)}]{Milosavljevic05b}
---. 2005b, arXiv:astro-ph/0512548

\bibitem[{{Niemiec} \& {Ostrowski}(2004)}]{Niemiec04}
{Niemiec}, J. \& {Ostrowski}, M. 2004, \apj, 610, 851

\bibitem[{{Nishikawa} {et~al.}(2003){Nishikawa}, {Hardee}, {Richardson},
  {Preece}, {Sol}, \& {Fishman}}]{Nishikawa03}
{Nishikawa}, K.-I., {Hardee}, P., {Richardson}, G., {Preece}, R.,
{Sol}, H., \&
  {Fishman}, G.~J. 2003, \apj, 595, 555

\bibitem[{{Ostrowski}(1991)}]{Ostrowski91}
{Ostrowski}, M. 1991, \mnras, 249, 551

\bibitem[{{Ostrowski}(1993)}]{Ostrowski93}
---. 1993, \mnras, 264, 248

\bibitem[{{Panaitescu} \& {Kumar}(2002)}]{Panaitescu02}
{Panaitescu}, A. \& {Kumar}, P. 2002, \apj, 571, 779

\bibitem[{{Peacock}(1981)}]{Peacock81}
{Peacock}, J.~A. 1981, \mnras, 196, 135

\bibitem[{Piran(2005)}]{Piran05}
Piran, T. 2005, AIP Conf. Proc., 784,
164,~arXiv:astro-ph/0503060

\bibitem[{{Romanov} {et~al.}(2004){Romanov}, {Bychenkov}, {Rozmus}, {Capjack},
  \& {Fedosejevs}}]{Romanov04}
{Romanov}, D.~V., {Bychenkov}, V.~Y., {Rozmus}, W., {Capjack},
C.~E., \&
  {Fedosejevs}, R. 2004, Physical Review Letters, 93, 215004

\bibitem[{{Silva} {et~al.}(2003){Silva}, {Fonseca}, {Tonge}, {Dawson}, {Mori},
  \& {Medvedev}}]{Silva03}
{Silva}, L.~O., {Fonseca}, R.~A., {Tonge}, J.~W., {Dawson}, J.~M.,
{Mori},
  W.~B., \& {Medvedev}, M.~V. 2003, \apjl, 596, L121

\bibitem[{Spitkovsky(2005)}]{Spitkovsky05}
Spitkovsky, A. 2005, AIP Conf. Proc., 801,
345,~arXiv:astro-ph/0603211

\bibitem[{{Vietri}(2003)}]{Vietri03}
{Vietri}, M. 2003, \apj, 591, 954

\bibitem[{{Vink} \& {Laming}(2003)}]{Vink03}
{Vink}, J. \& {Laming}, J.~M. 2003, \apj, 584, 758

\bibitem[{{V{\"o}lk} {et~al.}(2005){V{\"o}lk}, {Berezhko}, \&
  {Ksenofontov}}]{Volk05}
{V{\"o}lk}, H.~J., {Berezhko}, E.~G., \& {Ksenofontov}, L.~T. 2005,
\aap, 433,
  229

\bibitem[{{Wallace} \& {Epperlein}(1991)}]{Wallace91}
{Wallace}, J. M. \& {Epperlein}, E. M. 1991, Phys. Fluids B, 3,
1579

\bibitem[{{Wang} {et~al.}(2002){Wang}, {Loeb}, \& {Waxman}}]{Wang02}
{Wang}, X., {Loeb}, A., \& {Waxman}, E. 2002, \apj, 568, 830

\bibitem[Waxman \& Shvarts (1993)]{wax93}  Waxman, E., \& Shvarts, D.
  1993, Physics of Fluids A, 5, 1035

\bibitem[{{Waxman}(1995)}]{Waxman95}
Waxman, E. 1995, Phys. Rev. Lett., 75, 386

\bibitem[Waxman(1997)]{Waxman97} Waxman, E.\ 1997, \apjl, 485,
L5

\bibitem[{{Wiersma} \& {Achterberg}(2004)}]{Wiersma04}
{Wiersma}, J. \& {Achterberg}, A. 2004, \aap, 428, 365

\bibitem[Z\`{e}ldovich \& Raizer (1968)]{zel68}
  Z\`{e}ldovich, Ya. B. \& Raizer, Yu. P. 1968,
  Physics of Shock Waves and High-Temperature Hydrodynamic Phenomena, Chapter XII (New York:Academic).

\bibitem[{{Zhang}(1993)}]{Zhang93}
{Zhang}, M. 1993, Proc. 23th ICRC (Calgary) 2, 374

\bibitem[{{Zhang} \& {M{\'e}sz{\'a}ros}(2004)}]{Zhang04}
{Zhang}, B. \& {M{\'e}sz{\'a}ros}, P. 2004, International Journal of
Modern
  Physics A, 19, 2385
\end{thebibliography}

\end{document}